%% file: main.tex
\documentclass[prd,aps,letterpaper,twocolumn,preprintnumbers,superscriptaddress,nofootinbib,floatfix]{revtex4-1}

\usepackage{amsmath}
\usepackage{amssymb}
\usepackage{color}
\usepackage{graphicx}
\usepackage{dcolumn}
\usepackage{bm}
\usepackage[utf8]{inputenc}
\usepackage{xcolor}
\usepackage{tikz}
\usepackage{booktabs}
\AtBeginDocument{
\heavyrulewidth=.08em
\lightrulewidth=.05em
\cmidrulewidth=.03em
\belowrulesep=.65ex
\belowbottomsep=0pt
\aboverulesep=.4ex
\abovetopsep=0pt
\cmidrulesep=\doublerulesep
\cmidrulekern=.5em
\defaultaddspace=.5em
}

\usepackage[hypertexnames=false]{hyperref}
\hypersetup{
    colorlinks=true,       % false: boxed links; true: colored links
    linkcolor=blue,          % color of internal links
    citecolor=blue,        % color of links to bibliography
    filecolor=blue,      % color of file links
    urlcolor=blue           % color of external links
}

\DeclareMathOperator\tr{tr}
\newcommand{\avg}[1]{\left\langle #1 \right\rangle}
\newcommand{\obs}[0]{\mathcal{O}}
\newcommand{\defobs}[0]{\mathcal{Q}}

\DeclareMathOperator{\Var}{Var}
\let\Im\undefined
\let\Re\undefined
\DeclareMathOperator{\Re}{Re}
\DeclareMathOperator{\Im}{Im}
\DeclareMathOperator{\Tr}{Tr}

\begin{abstract}
Path integral contour deformations have been shown to mitigate sign and signal-to-noise problems associated with phase fluctuations in lattice field theories.
We define a family of contour deformations applicable to $SU(N)$ lattice gauge theory that can reduce sign and signal-to-noise problems associated with complex actions and complex observables. For observables, these contours can be used to define deformed observables with identical expectation value but different variance.
As a proof-of-principle, we apply machine learning techniques to optimize the deformed observables associated with Wilson loops in two dimensional $SU(2)$ and $SU(3)$ gauge theory. We study loops consisting of up to 64 plaquettes and achieve variance reduction of up to 4 orders of magnitude.
\end{abstract}

\begin{document}

\title{Path integral contour deformations for observables in \texorpdfstring{$SU(N)$}{SU(N)} gauge theory}

\author{William Detmold}
\affiliation{Center for Theoretical Physics, Massachusetts Institute of Technology, Cambridge, MA 02139, USA}
\affiliation{The NSF AI Institute for Artificial Intelligence and Fundamental Interactions}
\author{Gurtej Kanwar}
\affiliation{Center for Theoretical Physics, Massachusetts Institute of Technology, Cambridge, MA 02139, USA}
\affiliation{The NSF AI Institute for Artificial Intelligence and Fundamental Interactions}
\author{Henry Lamm}
\affiliation{Fermi National Accelerator Laboratory, Batavia, IL 60510, USA}
\author{Michael L. Wagman}
\affiliation{Fermi National Accelerator Laboratory, Batavia, IL 60510, USA}
\author{Neill C. Warrington}
\affiliation{Institute for Nuclear Theory, University of Washington, Seattle, Washington 98195-1550}
\preprint{FERMILAB-PUB-21-014-T}
\preprint{INT-PUB-21-002}
\preprint{MIT-CTP/5270}

\maketitle

\section{Introduction}  

In order to test the Standard Model and search for new physics in experiments involving hadrons and nuclei, precision calculations of Standard Model observables are required.
To achieve this, Monte Carlo (MC) calculations of lattice-regularized path integrals for quantum chromodynamics (QCD) have been used to make precise predictions for many phenomenologically relevant quantities in the meson and nucleon sectors; for recent reviews see Refs.~\cite{Aoki:2019cca,Detmold:2019ghl,Lehner:2019wvv,Cirigliano:2019jig,Aoyama:2020ynm}.
Lattice QCD calculations using MC methods are performed in Euclidean spacetime where the action $S(U)$ is typically a real function of the discretized gauge field $U_{x,\mu} \in SU(3)$ and an ensemble of gauge fields can be generated with probability distribution proportional to $e^{-S(U)}$.
Predictions for the expectation values of observables $\mathcal{O}$ are then obtained by averaging the results for $\mathcal{O}(U)$ obtained for each gauge field configuration.

For correlation functions describing nucleons, nuclei, and most mesons, $\mathcal{O}(U)$ is complex and includes a gauge-field-dependent phase~\cite{Wagman:2016bam,Wagman:2017gqi}.
Phase fluctuations grow more rapid with increasing Euclidean time separation and lead to a ``signal-to-noise (StN) problem'' in which the StN for a fixed size statistical ensemble decreases exponentially with increasing time separation~\cite{Parisi:1983ae,Lepage:1989hd,Beane:2009kya,Beane:2014oea,Wagman:2016bam,Wagman:2017gqi,Davoudi:2020ngi}.
Alternatively, multi-nucleon systems can be probed by including non-zero quark chemical potential in the action.
In this case, $e^{-S(U)}$ is not real and positive definite and cannot be interpreted as a probability distribution and the theory is said to have a ``sign problem.''
If $e^{-\Re S(U)}$ is used to define a probability distribution in this case, $e^{-i\Im S(U)}$ leads to rapid phase fluctuations of path integrands and the appearance of a StN problem that is exponential in the spacetime volume~\cite{Gibbs:1986ut,Cohen:2003kd,Cohen:2003ut,Splittorff:2006fu,Splittorff:2007ck,deForcrand:2010ys,Alexandru:2014hga}.

A generic method for taming sign and StN problems in path integrals has recently emerged. This method involves deforming the manifold of integration of the path integral into a complexified field space. If the path integrand is a holomorphic function of the field variables, then a multi-dimensional version of Cauchy’s integral theorem ensures that expectation values of the corresponding observable is unchanged by the manifold deformation. Manifold deformation may, however, change the values of observables on individual field configurations and therefore modify the severity of phase fluctuations and associated sign/StN problems. Several methods for finding useful manifolds have been proposed and successfully applied in lattice field theories as well as non-relativistic quantum mechanical theories relevant for condensed matter physics~\cite{Cristoforetti:2012su,Aarts:2013fpa,Cristoforetti:2013wha,Mukherjee:2013aga,Aarts:2014nxa,Cristoforetti:2014gsa,Alexandru:2015xva,Alexandru:2015sua,Alexandru:2016gsd,Fujii:2015vha,Tanizaki:2015rda,Alexandru:2017czx,Alexandru:2017lqr,Mori:2017nwj,Tanizaki:2017yow,Alexandru:2018brw,Alexandru:2018ngw,Alexandru:2018fqp,Alexandru:2018ddf,Kashiwa:2018vxr,Fukuma:2019uot,Fukuma:2019wbv,Kashiwa:2019lkv,Mou:2019gyl,Ulybyshev:2019fte,lawrence2020sign,Lawrence:2021izu}. For a recent review see Ref.~\cite{alexandru2020complex}.
Although most applications have focused on improving sign problems in theories with complex actions, an analogous method for improving StN problems for complex observables in theories with real actions was introduced in Ref.~\cite{Detmold:2020ncp}.

The focus of this work is on addressing sign/StN problems in QCD-like theories; in this setting, path integral contour deformations have so far been restricted to solving sign problems in simple contexts. Thermodynamic observables at non-zero quark chemical potential have been computed in $(0+1)$D QCD, a theory with a single link variable, using Lefschetz thimble methods~\cite{Di_Renzo_2018} and the generalized thimble method~\cite{Schmidt:2017gvu}, while preliminary results using neural-network manifolds were obtained in ~\cite{Ohnishi:2019ljc}. In higher dimensions, Lefschetz thimbles were used to analyze finite-density observables for one- and two-site systems in the heavy-dense limit in Ref.~\cite{zambello2018lefschetz}, and this study predicted that the number of relevant Lefschetz thimbles would grow exponentially with the number of lattice sites for larger systems.
The task of computing noisy observables in non-Abelian lattice gauge theories for larger spacetime volumes and higher spacetime dimensions has remained challenging.

This paper introduces a simple yet expressive family of complexified manifolds for taming sign/StN problems in $SU(N)$ gauge theory using path integral contour deformations.
The Jacobians required for calculations using this family of deformations are shown to be triangular matrices whose determinants can be computed with a cost that scales linearly with the spacetime volume.
This family of manifolds can be applied to all theories involving $SU(N)$ gauge fields, including gauge theories coupled to matter fields, although their practical utility for improving sign/StN problems is only explored here for pure gauge theory.

The deformed observable method introduced in Ref.~\cite{Detmold:2020ncp} relates path integrals over deformed contours to path integrals written in terms of modified observables on undeformed contours, enabling improvement in the StN of observables without the need to modify MC sampling.
We apply the method here to calculations of Wilson loops in $SU(2)$ and $SU(3)$ gauge theory, in which Wilson loops are known to have an exponentially severe StN problem and have been used to study other StN improvement methods~\cite{Luscher:2001up}.
Calculations are performed in $(1+1)$D as a proof of concept, as it is possible to compare with exact StN results derived analytically and to use specialized approaches for efficient Monte Carlo ensemble generation for $(1+1)$D gauge theories.
Results are obtained for a range of Wilson loop areas and lattice spacings including areas of up to 64 lattice units at the finest lattice spacing.
The variances of Wilson loops with largest areas are reduced by factors of $10^3$--$10^4$, demonstrating that deformed observables can dramatically improve StN problems in $SU(N)$ lattice gauge theory.
The linear scaling with spacetime volume of these contour deformations suggests that it should be computationally feasible to explore the application of analogous contour deformations to $(3+1)$D lattice gauge theory in future work.

The remainder of this paper is organized as follows.  Sec.~\ref{sec:gen} describes our approach to contour deformations for $SU(N)$ variables, including a family of complex manifolds for integration over sets of $SU(N)$ variables, and reviews the deformed observables method introduced in Ref.~\cite{Detmold:2020ncp}.  Sec.~\ref{sec:noise-suN} presents analytical results for expectation values and variances of observables in $(1+1)$D $SU(N)$ lattice gauge theory.
Results for MC calculations of deformed observables for Wilson loops are presented for $SU(2)$ gauge theory in Sec.~\ref{sec:su2} and for $SU(3)$ gauge theory in Sec.~\ref{sec:su3}. A summary of results and consideration of future work is found in Sec.~\ref{sec:cons}.

\section{General formalism}
\label{sec:gen}

Cauchy's integral theorem implies that the contour of a complex line integral can be deformed without changing the integral value if the integrand is holomorphic in the intervening region and the endpoints are held fixed.\footnote{For periodic functions this condition on the endpoints can be relaxed, as discussed in Sec.~\ref{sec:SUNdef}.}
When multidimensional integration is performed, the full theorem can be generalized if the integral is describable as iterated complex line integrals or by a technical extension to the full multivariate setting~\cite{range2013holomorphic}. For the purpose of contour deformations, however, only a weaker form of the theorem (equivalent to Stokes' theorem) is required. Specifically, a contour deformation from manifold $\mathcal{M}_A$ to $\mathcal{M}_B$ leaves the integral value unchanged if $\mathcal{M}_A \cup \mathcal{M}_B$ bounds a region in which the integrand is holomorphic; see Ref.~\cite{alexandru2020complex} for a simple proof.
To implement such contour deformations and confirm holomorphy of an integrand throughout the relevant region of configuration space, a coordinate parameterization is useful. We discuss such parameterizations and contour deformation for $SU(N)$ groups and $SU(N)$ gauge theory in the following sections.

\subsection{Contour deformations of angular parameters}\label{sec:SUNdef}

A general formalism for applying path integral contour deformations to $SU(N)$ group integrals can be obtained by using manifold coordinates that map subsets of $\mathbb{R}^{N^2-1}$ to $SU(N)$.
For any $N$, the group manifold can be given explicit global coordinates using $N^2-1$ angular variables~\cite{Bronzan:1988wa}. These variables can be divided into \emph{azimuthal} angles $\phi_1, \dots, \phi_J \in [0,2\pi]$ and \emph{zenith} angles $\theta_1, \dots, \theta_K \in [0, \pi/2]$, where $J = (N^2 + N - 2)/2$ and $K = (N^2 - N)/2$.\footnote{This is not the only possible assignment of angular coordinates to the manifold. For example, Appendix~\ref{app:su2-euler} explores an alternative parameterization for $SU(2)$.}
The azimuthal angles are periodic, such that $\phi_i = 0$ is identified with $\phi_i = 2\pi$, while the zenith angles have distinct endpoints. We define the combined coordinate $\Omega \equiv (\phi_1, \dots, \phi_J, \theta_1, \dots, \theta_K)$.

\begin{figure}
    \centering
    \input{contour1_cartoon}
    \caption{Left: schematic depiction of valid and invalid contour deformations, defined by the mapping $\widetilde{\theta}(\theta)$ from base coordinates to the manifold, when the original domain is a finite interval. Right: schematic depiction of additional allowed deformations (shifts) when endpoints are identified; these shifts are applicable to $U(1)$ variables or azimuthal angles $\phi$ in $SU(N)$ manifolds.} \label{fig:group_angle_deform}
\end{figure}
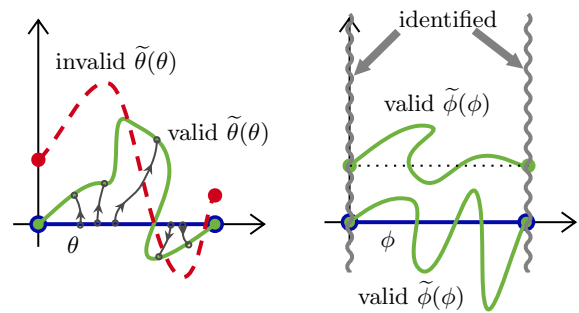

A generic integral over group-valued variable $U \in SU(N)$ can be written as
\begin{equation}
    \mathcal{I} = \int dU \, f(U),
\end{equation}
where the Haar measure $dU$ is defined to be the unique measure that satisfies $d(V U) = d( U V) = d U$ for $V \in SU(N)$. We choose the conventional normalization condition $\int d U = 1$.
Using the coordinates introduced above, the integral can immediately be cast as integration over a sub-domain of $\mathbb{R}^{N^2-1}$,
\begin{equation} \label{eq:sun-coord-integral}
    \mathcal{I} = \prod_{j=1}^{J} \left[
    \int_{0}^{2\pi} d\phi_j \right] \prod_{k=1}^{K} \left[ \int_{0}^{\pi/2} d\theta_k \right] h(\Omega) f(U(\Omega)),
\end{equation}
where $h(\Omega)$ is the Jacobian factor associated with the change of measure from $dU$ to $d\Omega = \prod_j d\phi_j \prod_k d\theta_k$, and $U(\Omega)$ is the group element at the manifold coordinate $\Omega$. The specific forms of $h(\Omega)$ and $U(\Omega)$ for the groups $SU(2)$ and $SU(3)$ are presented in Sec.~\ref{sec:su2-param} and \ref{sec:su3-param}.

From Eq.~\eqref{eq:sun-coord-integral}, it is clear that each $\theta_k$ can be deformed into the complex plane holding the endpoints $0$ and $\pi/2$ fixed. Each $\phi_j$ can be deformed under the weaker constraint that the endpoints remain identified, as shown in Fig.~\ref{fig:group_angle_deform}. This can be seen by noting that the endpoints of the shifted contour can be connected to the original endpoints using a pair of segments parallel to the imaginary axis; these segments differ by a real $2\pi$ shift and a change of orientation so that integrals of periodic functions along these segments exactly cancel. This approach to deforming periodic variables with identified endpoints has previously been applied to path integrals involving $U(1)$ variables~\cite{PhysRevD.97.094510,Detmold:2020ncp,Kashiwa:2020brj,Alexandru:2016ejd}. If the integrand $h(\Omega) f(U(\Omega))$ is a holomorphic function of all components of $\Omega$, the value of the integral is unchanged by the deformations described above. We can define a deformed integration contour by the maps 
$\widetilde{\phi}_j = \widetilde{\phi}_j(\Omega) \in \mathbb{C}$ and $\widetilde{\theta}_i=\widetilde{\theta}_i(\Omega) \in \mathbb{C}$;
for conciseness the collective set of deformed coordinates can be written as a function of original coordinates $\widetilde{\Omega} \equiv (\widetilde{\phi}_1, \dots, \widetilde{\phi}_J, \widetilde{\theta}_1, \dots, \widetilde{\theta}_K) = \widetilde{\Omega}(\Omega)$. The value of the integral is unchanged by this deformation,
\begin{widetext}
\begin{equation} \label{eq:deformed_group_integral}
\begin{aligned}
    \mathcal{I} &= \prod_{j=1}^{J} \left[
    \int_{0}^{2\pi} d\phi_j \right] \prod_{k=1}^{K} \left[ \int_{0}^{\pi/2} d\theta_k \right] J(\Omega) \, h(\widetilde{\Omega}) \, f(U(\widetilde{\Omega})) \\
    &= \prod_{j=1}^{J} \left[
    \int_{0}^{2\pi} d\phi_j \right] \prod_{k=1}^{K} \left[ \int_{0}^{\pi/2} d\theta_k \right] h(\Omega) \left\{ J(\Omega) \frac{h(\widetilde{\Omega})}{h(\Omega)} f(U(\widetilde{\Omega})) \right\} \\
    &= \int dU \, J(U) \, f(\widetilde{U}).
\end{aligned}
\end{equation}
\end{widetext}
Above, $\widetilde{U} \equiv U(\widetilde{\Omega})$ is the deformed group element, $J(\Omega) \equiv \det \frac{\partial \widetilde{\Omega}_\alpha}{\partial \Omega_{\beta}}$ is the (complex) Jacobian factor relating the measure $d\Omega$ of the base contour to the measure $d\widetilde{\Omega}$ of the deformed contour, and the Jacobian factor relating the Haar measure $dU$ between the original and deformed contours is given by
\begin{equation} \label{eq:deformed-jac-haar}
    J(U) \equiv J(\Omega) h(\widetilde{\Omega})/h(\Omega).
\end{equation}

For any concrete map $U(\Omega)$, the deformed group element is given by simply applying the map to the complexified coordinate $\widetilde{\Omega}$. Any real Lie group has a unique complexification and the space in which $\widetilde{U}$ lives is well understood. In particular, the complexification of $SU(N)$ is $SL(N,\mathbb{C})$~\cite{Bump2013}. This is easy to see on intuitive grounds, as $SU(N)$ matrices are specified by unit-norm eigenvalues with determinant $1$ and the effect of complexifying the group is to allow arbitrary non-zero eigenvalues while preserving the determinant constraint, resulting in the group $SL(N,\mathbb{C})$.

Though the deformation is defined in terms of the manifold coordinates, we can see in the last line of Eq.~\eqref{eq:deformed_group_integral} that this new integral can be written independently of the coordinates, using the modified integrand $J(U) f(\widetilde{U}(U))$.  This form suggests a coordinate-independent definition of a holomorphic integrand and contour deformation. Such a general approach is beyond the scope of this work, but we note that a deformation can be applied in any coordinate system so long as the integrand can be shown to be holomorphic in \emph{some} coordinate system. The angular coordinates for $SU(N)$ are sufficient to show that all components of the matrix representation of $U$ are holomorphic (i.e.~the Wirtinger derivatives $\partial U(\Omega) / \partial \bar{\Omega}_i$ are zero~\cite{range2013holomorphic}). The components of $U^{-1}$ are holomorphic whenever $U$ is invertible, as can be seen by starting with the definition of the inverse, $U^{-1} U = 1$, applying Wirtinger derivatives to both sides, $\partial(U^{-1} U)/\bar{\Omega}_i = 0$, and using holomorphy of $U$ to obtain $\partial U^{-1}/\partial \bar{\Omega}_i = 0$.

The general construction so far is valid for collections of $SU(N)$ group variables, and is therefore applicable to $SU(N)$ lattice gauge theory. Standard pure-gauge actions for $SU(N)$ lattice gauge theory can be written as holomorphic functions of the variables $U$ and their inverses, if we replace instances of $U^\dagger$ with $U^{-1}$, and are thus holomorphic throughout the $SL(N,\mathbb{C})$ domain of each complexified group variable. This fact has previously been recognized in complex Langevin approaches to extensive sign problems in lattice gauge theory~\cite{Aarts:2008rr,Sexty:2014zya,Seiler:2017wvd}. Similarly, (unrooted) fermionic determinants can be written as polynomials in components of gauge links and are therefore holomorphic~\cite{Alexandru:2018ngw}, and many observables can be analyzed in the same way.

\subsection{Vertical contour deformations}

We define a family of deformed manifolds for an $SU(N)$ variable in terms of the angular parameterization by
\begin{equation} \label{eq:vertical-deform}
\widetilde{\Omega} = \Omega + i f(\Omega; \lambda,\chi),
\end{equation}
where $f(\Omega; \lambda,\chi) \in \mathbb{R}^{N^2-1}$. This family of \emph{vertical deformations} inspired by Refs.~\cite{PhysRevD.97.094510, PhysRevLett.121.191602} is not fully general as it only includes contour deformations describable by vertically shifting each point purely in the imaginary direction, and in particular it does not include any deformations with $\Re{\widetilde{\Omega}(\Omega)} = \Re{\widetilde{\Omega}(\Omega')}$ for some $\Omega \neq \Omega'$. The Jacobian of any deformation in this family is straightforward to compute as
\begin{equation}
    J(\Omega) = \det \frac{\partial \widetilde{\Omega}_\alpha}{\partial \Omega_{\beta}} =  \det \left[ \delta_{\alpha\beta} + i \frac{\partial f_\alpha(\Omega;\lambda,\chi)}{\partial \Omega_{\beta}} \right],
\end{equation}
where $\alpha,\beta= 1,\ldots,N^2-1$ index the angular parameters $\Omega = (\phi_1,\ldots,\phi_J,\theta_1,\ldots,\theta_K)$.

The function $f$ can further be expanded in terms of Fourier modes,
\begin{equation} \label{eq:vertical-fourier-deform}
f(\Omega; \lambda, \chi) = \sum_{\{n_i\} = 0}^{\Lambda} \sum_{\{m_j\} = 1}^{\Lambda} \lambda_{I} T_{I}(\Omega; \chi_{I}^{1}, \dots, \chi_{I}^{a}),
\end{equation}
where $I \equiv (n_1 \dots n_a, m_1 \dots m_b)$, and
\begin{equation} \label{eq:vertical-fourier-deform2}
    T_{I}(\Omega; \chi^1 \dots \chi^a) \equiv \prod_{i=1}^{a}{\text{sin}(\phi_i n_i + \chi^i)} \prod_{j=1}^{b}{\text{sin}(2 \theta_j m_j)},
\end{equation}
which provides a complete basis for vertical contour deformations in the limit $\Lambda \rightarrow \infty$.
Including successively more Fourier modes by increasing the Fourier cutoff $\Lambda$ systematically improves the flexibility of the function $f$.
In our applications to $SU(N)$ gauge theories in $(1+1)$D below, Fourier cutoffs $\Lambda \in \{0,1,2\}$ are explored.
It is noteworthy that the sum over azimuthal Fourier modes includes the constant mode as well as phase offsets $\chi_i$ because azimuthal angles can be deformed without fixing their endpoints as discussed above.
These constant modes are essential for the sign/StN problem reduction achieved in $(1+1)$D examples below.

\subsection{Path integral deformations for noisy observables} \label{sec:obs-deforms}

We next review the deformed observable approach presented in Ref~\cite{Detmold:2020ncp}, in which contour deformations of lattice field theory path integrals are used to define modified observables with improved noise properties and unchanged expectation value. We focus here on $SU(N)$ lattice gauge theory path integrals, which are high-dimensional integrals over a collection of group-valued degrees of freedom $U_{x,\mu} \in SU(N)$. Here, $x$ specifies a site on the discrete spacetime lattice and $\mu \in \{1,\dots,N_d\}$ is any of the $N_d$ spacetime directions on the lattice. The integrals under study take the form 
\begin{equation}
  \begin{split}
      \avg{\mathcal{O}} \equiv \frac{1}{Z} \int \mathcal{D}U \ \mathcal{O}(U)\ e^{-S(U)},
  \end{split}\label{eq:Oave}
\end{equation}
where
\begin{equation}
\begin{split}
  Z = \int \mathcal{D} U\ e^{-S(U)}
\end{split}\label{eq:Z}
\end{equation}
and the action $S(U) \in \mathbb{R}$ is a function of all gauge links. Details on the construction of lattice gauge theory are presented in Sec.~\ref{sec:noise-suN}.

It is possible to deform the integration contour of an $SU(N)$ lattice gauge theory path integral by individually deforming each group-valued variable $U_{x,\mu}$ using the formalism presented above. In principle the deformed link, $\widetilde{U}_{x,\mu}$, could be a function of all other links on the lattice. However, evaluating the Jacobian factor arising from such an arbitrary deformation would require $O(V^3)$ operations, where $V$ is the number of sites of the lattice. For state-of-the-art lattice field theory calculations, this is intractable. A similar obstacle is encountered in the application of normalizing flows to sampling probability distributions in image or lattice data analysis, see e.g.\ Ref.~\cite{papamakarios2019normalizing} for a review, where it is avoided by explicitly restricting to triangular Jacobians for which the determinant factor is efficiently calculable from the diagonal elements. In this work, we similarly restrict to triangular Jacobians by allowing deformations of each variable to depend only on previous variables in a canonical ordering, described in detail for our particular studies of $SU(2)$ and $SU(3)$ in the following sections. Exploring other options is an interesting possibility for future work; for example, transformations built from a composition of multiple triangular transformations may allow more general spacetime dependence without significant increase in cost, whereas other alternatives based on convolutions may scale supra-linearly as the volume is increased.

Given a deformed manifold $\mathcal{M}$ with tractable Jacobian factor, a deformed integral can be constructed to compute the same  expectation value defined by Eq.~\eqref{eq:Oave},
\begin{equation} \label{eq:Oave-manif}
    \avg{\obs} = \frac{1}{Z} \int_{\mathcal{M}} \mathcal{D}\widetilde{U} \ \mathcal{O}(\widetilde{U})\ e^{-S(\widetilde{U})}.
\end{equation}
The above equality holds if the original manifold can be deformed to $\mathcal{M}$ without encountering any non-holomorphic regions of the integrand, i.e.~if the integrand is holomorphic in the region bounded by the union of $\mathcal{M}$ and the original manifold. The abstract manifold $\mathcal{M}$ can be specified by a map $\widetilde{U}(U)$, which then gives a concrete prescription for computing Eq.~\eqref{eq:Oave-manif},
\begin{equation} \label{eq:Oave-manif-concrete}
\begin{split}
    \avg{\obs} &= \frac{1}{Z} \int \mathcal{D}U \, J(U) \obs(\widetilde{U}(U)) \, e^{-S(\widetilde{U}(U))}.
\end{split}
\end{equation}
In contrast to Eq.~\eqref{eq:Oave}, the path integral in Eq.~\eqref{eq:Oave-manif-concrete} involves a generically complex-valued action, $S(\widetilde{U}(U)) \in \mathbb{C}$, and a different observable $J(U) \obs(\widetilde{U}(U))$ written in terms of the (efficiently computed) Jacobian factor $J(U)$. Cauchy's theorem implies that these modifications conspire to cancel and result in an identical expectation value.

It is useful to further rewrite Eq.~\eqref{eq:Oave-manif-concrete} as a path integral with respect to the original action,
\begin{equation} \label{eq:Qave}
\begin{split}
    \avg{\obs} &= \frac{1}{Z} \int \mathcal{D}U \left\{ J(U) \obs(\widetilde{U}(U)) e^{-S(\widetilde{U}(U)) + S(U)} \right\} e^{-S(U)} \\
    &\equiv \frac{1}{Z} \int \mathcal{D}U \, \defobs(U) \, e^{-S(U)} = \avg{\defobs}.
\end{split}
\end{equation}
In this rewriting, it is clear that the deformed path integral is still accessible by performing Monte Carlo sampling with respect to the original action $S(U)$ and estimating the sample mean of $\defobs(U)$. 
These methods can therefore be applied at the measurement step, after an ensemble of field configurations has been sampled. In essence, the deformed path integral defines a new observable $\defobs(U)$ that has provably identical expectation value to the original observable $\obs(U)$. Notably, this new observable generically has very different structure from $\obs$, as it may be non-local and depends on the structure of the action.

The variance of $\defobs(U)$ can, however, be vastly different from the variance of $\obs(U)$. For most observables, samples of $\obs(U)$ are complex-valued, and the variance of the real and imaginary components are given by
\begin{equation} \label{eq:Ovar}
\begin{split}
    \Var[\Re{\obs}] &= \frac{1}{2} \avg{|\obs^2|} + \frac{1}{2} \avg{\obs^2} - [\Re \avg{\obs}]^2, \\
    \Var[\Im{\obs}] &= \frac{1}{2} \avg{|\obs^2|} - \frac{1}{2} \avg{\obs^2} - [\Im \avg{\obs}]^2.
\end{split}
\end{equation}
These are not generically identical to the variance of $\Re{\defobs}$ and $\Im{\defobs}$,
\begin{equation} \label{eq:Qvar}
\begin{split}
    \Var[\Re{\defobs}] &= \frac{1}{2} \avg{|\defobs^2|} + \frac{1}{2} \avg{\defobs^2} - [\Re \avg{\defobs}]^2, \\
    \Var[\Im{\defobs}] &= \frac{1}{2} \avg{|\defobs^2|} - \frac{1}{2} \avg{\defobs^2} - [\Im \avg{\defobs}]^2.
\end{split}
\end{equation}
The final term is unchanged because $\avg{\obs} = \avg{\defobs}$, but the first two terms in both lines of Eq.~\eqref{eq:Qvar} are not generically equal to the corresponding terms in Eq.~\eqref{eq:Ovar}. Explicitly, those terms are given by
\begin{equation} \label{eq:var-non-holo-terms}
\begin{split}
    \avg{|\defobs^2|} &= \frac{1}{Z} \int \mathcal{D}U \left| J(U) \obs(\widetilde{U}(U)) \right|^2 e^{-2\Re{S(\widetilde{U}(U))} + S(U)}, \\
    \avg{\defobs^2} &= \frac{1}{Z} \int \mathcal{D}U J(U)^2 \obs(\widetilde{U}(U))^2 e^{-2 S(\widetilde{U}(U)) + S(U)}.
\end{split}
\end{equation}
Extra factors of $S(U)$ persist in both expressions in Eq.~\eqref{eq:var-non-holo-terms}, and a non-holomorphic absolute value appears for $\avg{|\defobs^2|}$, preventing identification with the terms in Eq.~\eqref{eq:Ovar}. It can thus be fruitful to look for a modified observable $\defobs$ for which the terms in Eq.~\eqref{eq:var-non-holo-terms} are minimized and the statistical noise is less than that of the original observable.

Given an explicit parameterization of the deformed contour, standard gradient-based optimization methods can be applied to find the parameters that minimize the terms in Eq.~\eqref{eq:var-non-holo-terms}. Since the parameters only affect the observable itself (the sampling weight is always $e^{-S(U)}$), the gradient of the variance with respect to the parameters can be written as an expectation value under the original Monte Carlo sampling. In this work, minimization of Eq.~\eqref{eq:var-non-holo-terms} is performed using stochastic gradient descent over the deformed contour parameters, which approximately converges to a local minimum of the variance under mild assumptions~\cite{robbins1951,borkar2009stochastic,chee2018convergence,bottou2018optimization}, and starting at the original manifold ensures that the result improves relative to (or at worst is equivalent to) the original variance.

A potential obstacle to the deformed observable method is that some contour deformations could lead to a severe overlap problem between probability distributions proportional to $e^{-S(U)}$ and $e^{-\text{Re}[S(\widetilde{U}(U))]}$ that  could make estimates of deformed observable variances from finite Monte Carlo ensembles unreliable. The possibility of constructing deformed observables with severe overlap problems can be mitigated, however, by constructing deformed observables that minimize the variance of $\mathcal{Q}$ since large fluctuations of $e^{-S(\widetilde{U}(U))+S(U)}$ will tend to increase the variance of $\mathcal{Q}$. Overlap problems could still arise due to overfitting to the sample variance of a Monte Carlo ensemble that does not sample the field configurations associated with large fluctuations of $e^{-S(\widetilde{U}(U))+S(U)}$; practical strategies to avoid such overfitting problems are discussed in Sec.~\ref{sec:su2-training} below.

The terms to minimize in Eq.~\eqref{eq:var-non-holo-terms} are specific to a particular observable $\obs$. There is no reason to expect a single manifold deformation to be optimal for all possible observables, but one does expect similar observables to be highly correlated, and thus to receive similar variance improvements from the same deformation. This suggests two useful practical improvements:
\begin{enumerate}
    \item Optimal manifolds can be found for a few representative observables and can be reused for other similar observables.
    \item When optimizing manifold parameters, the optimal parameters for a similar observable can be used to initialize the search.
\end{enumerate}
In our studies of $SU(2)$ and $SU(3)$ lattice gauge theory in Sec.~\ref{sec:su2} and Sec.~\ref{sec:su3}, the initialization approach was found to significantly reduce the number of steps required for optimization.

\subsection{Rewriting observables before deformation}\label{sec:rewriting}

It is often the case that the expectation value of an observable $\mathcal{O}$ has multiple equivalent path integral representations, for example when a theory possesses a gauge or global symmetry that modifies $\mathcal{O}$ but leaves the action and expectation values of observables invariant.
In addition to the freedom of choosing the parameterization of contour deformations for a given path integral discussed above, the application of contour deformations to observables includes the freedom of choosing which path integral representation to deform.

\begin{figure*}
    \centering
    \includegraphics{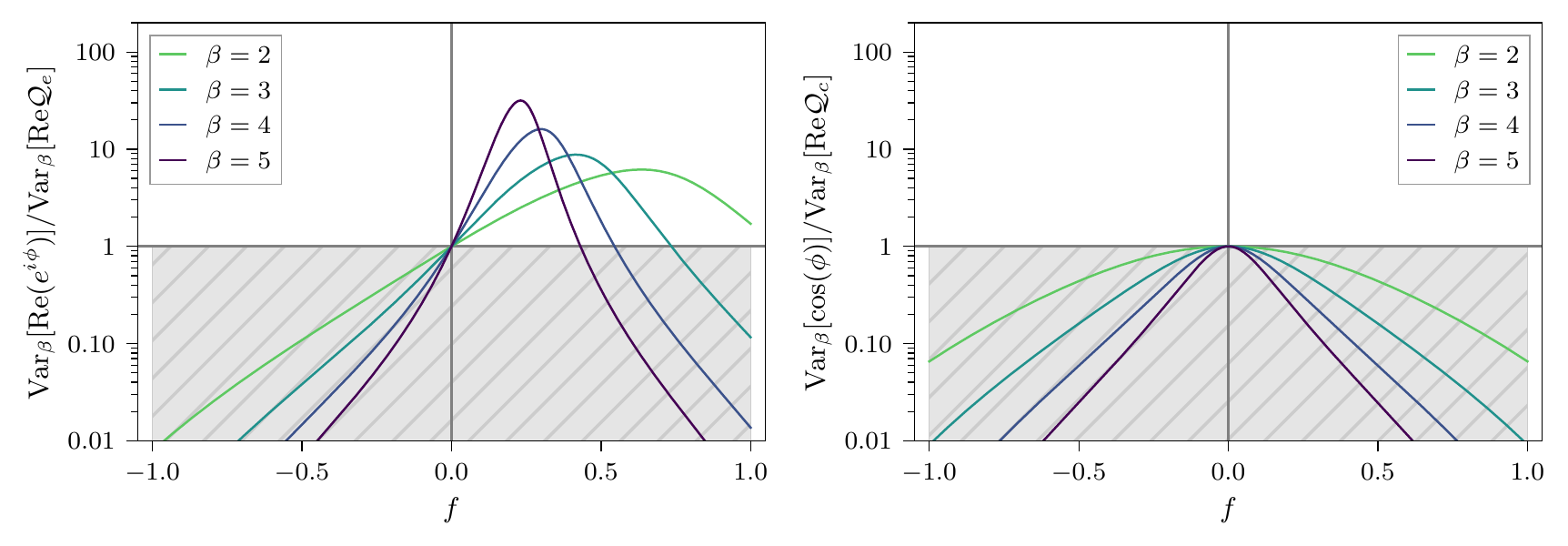}
    \caption{Ratios of the variance of observable $\Re(e^{i\phi}) = \cos(\phi)$ to the variance of deformed observables obtained by applying the transformation $\phi\rightarrow \phi + i f$ to the integrand $e^{i\phi}$ (left) or $\cos(\phi)$ (right) in the one-dimensional path integral with Euclidean action $-\beta \cos\phi$ discussed in the main text. Gray hatching indicates the region in which the variance of the deformed observable is higher than the original (i.e.~there is no improvement). \label{fig:1site}}
\end{figure*}

Simple examples of this freedom arise in two-dimensional $U(1)$ Euclidean lattice gauge theory. For instance, path integral representations of Wilson loops with unit area in this theory are proportional to
\begin{equation}
   \int_{-\pi}^{\pi} \frac{d\phi}{2\pi}\ e^{i\phi}\ e^{\beta \cos \phi} = I_1(\beta),
    \label{eq:I1expi}
\end{equation}
which is an integral representation of the modified Bessel function of the first kind $I_n(\beta)$, with $n=1$, written in terms of the field variable $\phi$.
The integrand appearing can be interpreted as a product of an observable $\obs = e^{i\phi}$ and an $e^{-S}$ factor for the Euclidean action $S = -\beta\cos\phi$.
This theory has a charge conjugation symmetry that acts on the integrand of Eq.~\eqref{eq:I1expi} by $\phi \rightarrow -\phi$, which leaves the action invariant but modifies the observable $e^{i\phi} \rightarrow e^{-i\phi}$.
An equally valid integral representation of $I_1(\beta)$ is obtained by averaging the observable and its charge conjugate as 
\begin{equation}
    I_1(\beta) = \int_{-\pi}^{\pi} \frac{d\phi}{2\pi}\ \cos(\phi)\ e^{\beta \cos \phi}.
    \label{eq:I1cos}
\end{equation}
The choice of integrand in Eq.~\eqref{eq:I1expi} or Eq.~\eqref{eq:I1cos} is irrelevant for Monte Carlo calculations using the integration contours shown because the Monte Carlo estimator used in the first case is $\Re{e^{i\phi}} = \cos{\phi}$.
However, the ability of path integral contour deformations to reduce the variance of a Monte Carlo calculation does depend on the representation. In particular, the variance of a Monte Carlo evaluation of Eq.~\eqref{eq:I1expi} can be significantly reduced by contour deformations while the variance of a Monte Carlo evaluation of Eq.~\eqref{eq:I1cos} cannot, as discussed below and shown in Fig.~\ref{fig:1site}.

Ref.~\cite{Detmold:2020ncp} demonstrated that the variance of two-dimensional $U(1)$ Wilson loops can be significantly reduced using contour deformations of the representation given in Eq.~\eqref{eq:I1expi}; we review the analytical derivation here.
Denote averaging of $\mathcal{O}(\phi)$ with respect to the path integral of the theory with action $-\beta \cos{\phi}$ by 
\begin{equation}
    \left<\mathcal{O}(\phi)\right>_\beta \equiv \int_{-\pi}^{\pi} \frac{d\phi}{2\pi I_0(\beta)} \mathcal{O}(\phi) e^{\beta \cos{\phi}},
\end{equation}
where the modified Bessel function with rank $n=0$ is used to normalize the distribution. In general, the modified Bessel functions of the first kind are given by 
$I_n(\beta) = \int_{-\pi}^{\pi} \frac{d\phi}{2\pi} e^{i n \phi} e^{\beta \cos\phi}$
and will be used throughout the following derivation.
Further denoting the corresponding variance of $\mathcal{O}$ given $\beta$ by $\text{Var}_\beta[\mathcal{O}]$, the variance of $\Re{e^{i\phi}} = \cos\phi$ can be computed to be
\begin{equation}
\begin{split}
    \text{Var}_\beta[\Re{e^{i\phi}}] 
    &= \left<\cos^2(\phi)\right>_\beta - \left<e^{i\phi}\right>_\beta^2 \\
    &= \frac{1}{2}\left[1 + \frac{I_2(\beta)}{I_0(\beta)}\right] - \left[\frac{I_1(\beta)}{I_0(\beta)}\right]^2.
    \end{split}
\end{equation}
The constant vertical deformation 
\begin{equation}
    \phi \; \rightarrow \; \widetilde{\phi}(\phi) = \phi + i f
    \label{eq:constantshift}
\end{equation}
leads to a deformed observable 
\begin{equation}
    \mathcal{Q}_e(\phi) = e^{i\widetilde{\phi}(\phi)} e^{\beta\cos(\widetilde{\phi}(\phi)) - \beta\cos(\phi)}
\end{equation}
satisfying $\left<\mathcal{Q}_e\right>_\beta = \left< e^{i\phi} \right>_\beta$. Using $\Re{\mathcal{Q}_e}$ as a Monte Carlo estimator instead of $\Re{e^{i\phi}}$ results in variance
\begin{equation}
\begin{split}
    \text{Var}_\beta[\Re{\mathcal{Q}_e} ]
    &= e^{-2 f} \left[ \frac{R(\beta,f)}{2} +  V(\beta,f)  \right] - \left[\frac{I_1(\beta)}{I_0(\beta)}\right]^2,
    \end{split}
            \label{eq:1sitevarexpi}
\end{equation}
where
\begin{equation}
\begin{split}
   V(\beta,f) &= \left(\frac{e^{f} - \frac{1}{2}}{e^{-f} - \frac{1}{2}}\right)\frac{I_2(\beta\sqrt{5 - 4\cosh(f)})}{2 I_0(\beta)}, \\
  R(\beta,f) &=
   \frac{I_0(\beta(2 \cosh (f) - 1))}{I_0(\beta)}.
\end{split}
\end{equation}
For positive $\beta$, $\text{Var}_\beta[\Re{\mathcal{Q}_e}]$ generically has a minimum at some strictly positive $f$ that defines the optimal contour for reducing the variance of $\mathcal{Q}_e$.
Using deformed observables associated with this optimal contour rather than the original contour leads to variance reduction whose significance is larger than an order of magnitude for $\beta \gtrsim 4$ as shown in Fig.~\ref{fig:1site}.
Larger variance reductions can be obtained for multi-dimensional generalizations of Eq.~\eqref{eq:I1expi} associated with larger $U(1)$ Wilson loops~\cite{Detmold:2020ncp}.

Applying the same constant vertical deformation to the representation in Eq.~\eqref{eq:I1cos} leads to an alternative deformed observable
\begin{equation}
    \mathcal{Q}_c(\phi) = \cos(\widetilde{\phi}(\phi))\ e^{\beta\cos(\widetilde{\phi}(\phi)) - \beta\cos(\phi)} 
\end{equation}
satisfying $\left<\mathcal{Q}_c\right>_\beta = \left< e^{i\phi} \right>_\beta$.
The real part of this alternative deformed observable has variance
\begin{equation}
\begin{split}
    \text{Var}_\beta[\Re{\mathcal{Q}_c} ] 
    &= \frac{1}{2}e^{-2f} V(\beta,f) + \frac{1}{2}e^{2f} V(\beta,-f)  \\
      &\hspace{10pt} + \frac{1}{2}R(\beta,f)  - \left[\frac{I_1(\beta)}{I_0(\beta)}\right]^2.
      \end{split}
\end{equation}
This expression is symmetric under $f \rightarrow -f$ and its gradient with respect to $f$ therefore vanishes at $f=0$, which can be verified to be the minimum of $f$. The deformed observable $\mathcal{Q}_c$ associated with integrating along the real axis is thus the choice with lowest variance, while increasing $|f|$ away from zero always leads to increased variance as illustrated for several choices of $\beta$ in Fig.~\ref{fig:1site}.

The qualitative features of these results can be understood from the behaviors of magnitude and phase fluctuations of deformed observables.
The magnitude of $e^{i \widetilde{\phi}}$ is reduced for all $\phi$ by a constant vertical deformation with $f > 0$. To preserve the results of the holomorphic integral in Eq.~\eqref{eq:I1expi} under this deformation, cancellations from phase fluctuations must correspondingly be less severe and sign/StN problems should therefore be improved.
On the other hand, the magnitude of $\cos(\widetilde{\phi})$ always satisfies $|\cos(\widetilde{\phi})| \geq |\cos(\phi)|$ and therefore can only be increased by applying a vertical contour deformation.
This suggests that phase fluctuations must become more severe to preserve deformation-invariant integral results and that sign/StN problems will be worsened.
This argument applies to non-constant vertical deformations and suggests that it is generically difficult to construct a contour deformation of Eq.~\eqref{eq:I1cos} that leads to a deformed observable with variance comparable to the observable $\mathcal{Q}_e$ obtained by deforming Eq.~\eqref{eq:I1expi}.

In the non-Abelian gauge theories that are focus of this work, the traces of Wilson loops define gauge invariant observables related to the potential energies of static quark configurations analogous to $e^{i\phi}$ in $U(1)$ gauge theory.
The eigenvalues of Wilson loops in $U(N)$ or $SU(N)$ gauge theories can be expressed as $e^{i\phi_j}$, where $\phi_j \in \mathbb{R}$ for $j=1,\ldots,N$ and for the case of $SU(N)$ the phases satisfy $\sum_{j=1}^N \phi_j = 0 \mod 2\pi$.
The trace of the Wilson loop is given by $\sum_j e^{i\phi_j}$.
For $SU(N)$ gauge theory, the unit determinant condition therefore results in analogous obstacles to improving the variance of the trace of Wilson loops if the observable is not first rewritten using symmetries.

For the case of $SU(2)$, there is a precise correspondence between Wilson loop traces and Eq.~\eqref{eq:I1cos}. The unit determinant condition requires the Wilson loop eigenvalues to be of the form $e^{i\phi}$ and $e^{-i\phi}$ and the trace appearing in both the observable and the Wilson action~\cite{Wilson:1974sk} (discussed below in Sec.~\ref{sec:su2}) are proportional to $\cos(\phi)$.
It is therefore similarly impossible to improve the variance of a unit area $SU(2)$ Wilson loop trace using constant vertical deformations, and the previous analysis suggests it is difficult to make significant variance improvements to traces of $SU(2)$ Wilson loops of general area using vertical contour deformations.
However, this analysis also indicates that $e^{i\phi}$ could provide a suitable starting point for defining deformed observables. The gauge invariant component of the Wilson loop eigenvalue $e^{i\phi}$ is $\cos(\phi)$ making this a suitable rewriting of the observable that does not affect the expectation value but enables variance improvements by contour deformation.\footnote{We could instead use linearity of the expectation value to start from an explicitly gauge-invariant observable and rewrite $\left< \cos(\phi) \right> = \frac{1}{2}\left< e^{i\phi} \right> + \frac{1}{2}\left< e^{-i\phi} \right>$. Most generally, we could then apply distinct deformations to each expectation value on the right-hand-side of this expression, and in particular applying constant imaginary shifts with opposite signs will result in the same StN improvement for both terms. In this example, the two expectation values can be equated using charge conjugation symmetry, which allows the exact rewriting $\left<\cos(\phi)\right> = \left<e^{i\phi}\right>$, but the technique of splitting the expectation value using linearity and applying independent deforms may be useful where such a symmetry is not present.}

In the case of $SU(N)$ with $N \geq 3$, complex eigenvalues are not guaranteed to come in complex conjugate pairs,
but the unit determinant condition still guarantees that any vertical deformation of the eigenvalue phases will increase the magnitude of at least one $e^{i\phi_j}$ phase factor. In the sum over eigenvalues defining the trace of a Wilson loop, the largest eigenvalue magnitude will set the typical observable magnitude on generic gauge fields in the integration domain. While it is possible for stronger cancellation between eigenvalues to occur on particular gauge fields, this larger typical magnitude throughout the majority of the domain of integration suggests that cancellations from phase fluctuations with similar severity to those of the original observable will therefore be required for such deformed observables to achieve identical expectation values, and significant sign/StN problems may be expected.
If one instead chooses the non-gauge-invariant integrand $e^{i\phi_1}$ to measure the same expectation value, the determinant constraint does not prevent exponentially decreasing the magnitude throughout the integration domain using contour deformations. For example, one can choose the shift \begin{equation}
\begin{aligned}
    \phi_1 \; &\rightarrow \; \widetilde{\phi}_1 = \phi_1 + i f \\
    \phi_2 \; &\rightarrow \; \widetilde{\phi}_2 = \phi_2 - i f / 2 \\
    \phi_3 \; &\rightarrow \; \widetilde{\phi}_3 = \phi_3 - i f / 2
\end{aligned}
\end{equation}
which has the effect of reducing the observable magnitude by $e^{-f}$ on every gauge field in the integration domain.

Rewriting Wilson loop observables based on eigenvalues requires diagonalization, and a more practical alternative is to use the $(1,1)$ component of the (matrix-valued) Wilson loop as another non-gauge-invariant function with the same expectation value as the trace divided by $N$.
The phase of this (or any other) single color component of the Wilson loop is not constrained by the unit determinant condition and therefore one expects that a suitable parameterization can be found in which vertical deformations can be applied to the phase of the $(1,1)$ component of the Wilson loop analogously to $e^{i\phi}$.
Such parameterizations are given for $SU(2)$ in Sec.~\ref{sec:su2} and for $SU(3)$ in Sec.~\ref{sec:su3} following Ref.~\cite{Bronzan:1988wa} and are used as starting points for defining deformed observables with reduced variance in calculations of Wilson loop expectation values.
An alternative parameterization of $SU(2)$ in which the real part of the $(1,1)$ component of the Wilson loop is expressed as $\cos(\alpha/2)$ is explored in Appendix~\ref{app:su2-euler}; as expected, vertical contour deformations do not improve the variance of unit area Wilson loops and less (though still significant) variance reduction is found for larger area Wilson loops with this alternative parameterization.

\section{Noise problems in \texorpdfstring{$SU(N)$}{SU(N)} lattice gauge theory}
\label{sec:noise-suN}
A simple setting for analyzing $SU(N)$ lattice gauge theory is obtained by considering $(1+1)$D Euclidean spacetime with open boundary conditions.
In this spacetime geometry, much like in $(3+1)$ dimensions, the theory features confinement of static test charges and an exponentially severe StN problem associated with static quark correlation functions, which can be identified with Wilson loops~\cite{Wilson:1974sk}.
Numerical calculations of Wilson loop expectation values can be performed at much lower computational cost in $(1+1)$D than $(3+1)$D,  facilitating a first exploration of path integral contour deformations applied to non-Abelian gauge theory observables on non-trivial lattices.
Analytic results for $(1+1)$D observables such as Wilson loops are also known~\cite{Wadia:2012fr,Gross:1980he} and can be used to verify the correctness of numerical results.
These results are extended to analytic results for the variances of $(1+1)$D Wilson loops below, which are then used in Secs.~\ref{sec:su2}--\ref{sec:su3} to verify the correctness and study the effectiveness of contour deformations applied to Wilson loops. In particular, analytic results can be used to determine the StN gains obtained by using deformed observables even when the corresponding undeformed observables are too noisy to be determined reliably.

\subsection{\texorpdfstring{$SU(N)$}{SU(N)} lattice gauge theory in \texorpdfstring{$(1+1)$D}{(1+1)D}} \label{sec:suN-defns}

Lattice gauge theory in $(1+1)$D is defined on a set $\mathcal{V}$ of Euclidean spacetime points $x$ arranged in a discrete two-dimensional lattice, with vectors $\hat{1}$ and $\hat{2}$ giving the displacement in lattice units between neighboring lattice sites along the two Euclidean spacetime axes.
The discretized gauge field is represented by group-valued variables on each link of the lattice, with $U_{x,\mu}$ denoting the variable associated with link $(x,x+\hat{\mu})$.
The physical content of the theory is encoded in the (discretized) action. We consider the Wilson action for $SU(N)$ lattice gauge theory~\cite{Wilson:1974sk}, given for a $(1+1)$D Euclidean spacetime volume by
\begin{equation}\label{eq:Wilson-Action-2d}
   S \equiv - \frac{1}{g^2} \sum_{x \in \mathcal{V}} \tr \left( P_{x} + P_x^{-1} \right) ,
\end{equation}
where $g$ is the bare gauge coupling and each \emph{plaquette} $P_x \in SU(N)$ is defined as
\begin{equation}
    P_x \equiv U_{x,1} U_{x+\hat{1},2} U^{-1}_{x+\hat{2},1} U^{-1}_{x,2}.
\end{equation}
Writing the action and plaquettes using inversion rather than Hermitian conjugation allows the relevant integrands to be interpreted in the following sections as holomorphic functions of integration variables throughout the complexified domain. For $SU(N)$ elements these operations are equivalent, but analytically continuing the action to $SL(N,\mathbb{C})$ requires the use of the inverse~\cite{Aarts:2008rr,Sexty:2014zya,Seiler:2017wvd}.

Expectation values of operators $\mathcal{O}(U)$ in the lattice regularized theory are defined by specializing Eq.~\eqref{eq:Oave}--\eqref{eq:Z} to the particular case of $SU(N)$ lattice gauge theory
\begin{equation}
  \begin{split}
      \avg{\mathcal{O}} = \frac{1}{Z} \int \mathcal{D}U \ \mathcal{O}(U)\ e^{-S(U)},
  \end{split}\label{eq:Oave2}
\end{equation}
where the Euclidean partition function $Z$ is defined by
\begin{equation}
\begin{split}
  Z = \int \mathcal{D} U\ e^{-S(U)}
\end{split}
\end{equation}
and $\mathcal{D}U = \prod_{x,\mu} d U_{x,\mu}$ in terms of the Haar measure $d U_{x,\mu}$ of $SU(N)$.

With open boundary conditions in $(1+1)$D, the partition function defined by this action factorizes into a product of independent integrals over each $P_x$.
To exploit this factorization in $(1+1)$D, a gauge fixing prescription can be applied in which $U_{x,2} = 1$ for all $x$ and $U_{x,1}=1$ for sites with $x_2 = 0$ (a maximal tree gauge). In this gauge, 
\begin{equation} \label{eq:plaq-to-link}
P_{x} = U_{x,1} U_{x+\hat{2},1}^{-1},
\end{equation}
which can be easily inverted to obtain
\begin{equation} \label{eq:link-to-plaq}
    U_{x,1} = \left[ \prod_{k=0}^{x_2-1} P_{x + k \hat{2}} \right]^{-1}.
\end{equation}
The variables $P_x$ are therefore in one-to-one correspondence with the remaining non-gauge-fixed $U_{x,1}$.
The Haar measure is invariant under this change of variables, and the path integral defining the partition function factorizes as
\begin{equation}
   \begin{split}
      Z = \prod_{x \in \mathcal{V}'} z = z^{|\mathcal{V}'|},
   \end{split}\label{eq:Zfac}
\end{equation}
where $\mathcal{V}' \subset \mathcal{V}$ is the subset of lattice points with unconstrained $U_{x,1}$ in this gauge (those for which $x_2 \neq 0$)
and $z$ is the contribution to the partition function from a single plaquette,
\begin{equation}
   \begin{split}
      z \equiv  \int d P \ e^{\frac{1}{g^2} \tr \left( P + P^{-1} \right)}.
   \end{split}\label{eq:zdef}
\end{equation}
The calculations of $z$ and similar single-variable $SU(N)$ integrals are presented in Appendix~\ref{app:single-site-integrals}.

Wilson loops are defined by the matrix-valued quantity
\begin{equation}
   \begin{split}
      W_{\mathcal{A}} \equiv \prod_{x,\mu \in \partial \mathcal{A}} U_{x,\mu},
   \end{split}\label{eq:Wdef}
\end{equation}
where $\prod_{x,\mu \in \partial \mathcal{A}} U_{x,\mu}$ represents an ordered product of links along the boundary $\partial \mathcal{A}$ of the two-dimensional region $\mathcal{A}$ with area $A$. The expectation value of the gauge-invariant observable $ \frac{1}{N} \tr \left( W_{\mathcal{A}} \right)$
probes the interaction between a pair of static quarks if the region $\mathcal{A}$ is taken to be rectangular.
Inserting Eq.~\eqref{eq:link-to-plaq} into Eq.~\eqref{eq:Wdef} gives\footnote{For simplicity we restrict to rectangular Wilson loops with one corner at the origin.}
\begin{equation}
   \begin{split}
      \frac{1}{N} \tr \left( W_{\mathcal{A}} \right) = \frac{1}{N} \tr \left( \prod_{x \in \mathcal{A}} P_x \right).
   \end{split}\label{eq:Wfac1}
\end{equation}
Using linearity of expectation values and factorization of path integrals analogous to Eq.~\eqref{eq:Zfac}, the expectation values of Wilson loops can be related to products of (matrix-valued) single-variable expectation values,
\begin{equation}
    \left< \frac{1}{N} \tr \left( W_{\mathcal{A}} \right) \right> = \frac{1}{N} \tr \left( \prod_{x\in\mathcal{A}} \avg{P_x} \right).
\end{equation}
Each single-variable expectation value is given by $\left< P_{x}^{ab} \right> = \avg{\chi_1}\delta^{ab}$, allowing the traced Wilson loop to be written as a product of scalars,
\begin{equation}
   \begin{split}
      \avg{ \frac{1}{N} \tr \left( W_{\mathcal{A}} \right) } &= \prod_{x \in \mathcal{A}} \avg{\chi_1} = \avg{\chi_1}^A, \\
   \end{split}\label{eq:Wfac}
\end{equation}
where we have introduced the single-variable normalized expectation value of the group character function $\chi_1(P)=\tr(P)$,
\begin{equation}
   \begin{split}
      \avg{\chi_1} &\equiv \frac{1}{z} \int dP \ \frac{1}{N}\  \tr (P) \ e^{\frac{1}{g^2} \tr \left( P + P^{-1} \right)},
   \end{split}\label{eq:wdef}
\end{equation}
whose value is computed in Appendix~\ref{app:single-site-integrals}.

Eq.~\eqref{eq:Wfac} implies that Wilson loop expectation values follow area law scaling, $\avg{ \tr ( W_{\mathcal{A}} ) / N } \sim e^{-\sigma A}$, and $SU(N)$ gauge theory in $(1+1)$D confines for all values of the coupling, with a separation-independent force between static test charges given by the string tension
\begin{equation}
      \sigma \equiv - \lim_{A\rightarrow \infty} \partial_A \ln W_{\mathcal{A}}
      = -\ln \avg{\chi_1}.\label{eq:sigmadef}
\end{equation}
Although $\avg{\chi_1}$ is in general given by a convergent infinite series in Eq.~\eqref{eq:avgchi}, in the case of $SU(2)$ a simpler form can be found in terms of modified Bessel functions,
\begin{equation}
   \begin{split}
      \sigma^{SU(2)} = \ln\left( \frac{I_1(4/g^2)}{I_2(4/g^2)} \right),
   \end{split}\label{eq:SU2sigma}
\end{equation}
which goes to zero as $g^2 \rightarrow 0$. This observation can be generalized to all $SU(N)$ groups, and the lattice-units string tension goes to zero while the static quark correlation length grows to infinity in the limit of $g^2 \rightarrow 0$ in all cases. We can consider this to be the naive continuum limit of the theory, though the correlation lengths of dynamical quantities such as plaquettes or localized Wilson loops remain finite by the factorization of the path integral. When investigating the approach to the continuum in Sec.~\ref{sec:su2} and \ref{sec:su3}, we should decrease the coupling while fixing the dimensionless quantity $\sigma V$, where $V$ is the total number of plaquettes; the particular choices of couplings and $V$ used in our numerical studies are reported in Table~\ref{tab:couplings}. Results are plotted versus $\sigma A$ when comparing quantities at fixed physical separation is important.

\begin{center}
\begin{table}[]
    \centering
    \begin{tabular}
    {>{\centering}m{0.75cm} >{\centering}m{0.75cm} @{\hspace{.5cm}} *{3}{>{\centering}m{1cm}} >{\centering\arraybackslash}m{1cm}}
    \toprule
    & & \multicolumn{2}{c}{$SU(2)$} & \multicolumn{2}{c}{$SU(3)$} \\
    \cmidrule(lr){3-4} \cmidrule(lr){5-6}
    $\sigma$ & $V$ & $g$ & $\beta$ & $g$ & $\beta$      \\
    \midrule
    0.4    & $16$ & 0.98 & 4.2 & 0.72 & 11.7\\
    0.2    & $32$ & 0.71 & 8.0 & 0.53 & 21.7\\
    0.1    & $64$ & 0.51 & 15.5 & 0.38 & 41.8\\
    \bottomrule
    \end{tabular}
    \caption{The couplings used in our numerical studies of $SU(2)$ and $SU(3)$ lattice gauge theory. The dimensionless quantity $\sigma V$ is fixed to $6.4$ while $\sigma$ and $V$ are individually varied. The conventional Wilson action parameter $\beta = 2N / g^2$ is also reported.}
    \label{tab:couplings}
\end{table}
\end{center}

\subsection{Noise and sign problems in the Wilson loop}

Although the expectation value  $\avg{ \tr ( W_{\mathcal{A}} ) / N }$ is real, the integrand $\tr ( W_{\mathcal{A}} ) / N$ has fluctuating signs (for $N = 2$) or fluctuating complex phases (for $N \geq 3$) across the domain of integration. These fluctuations result in a sign/StN problem for this observable. The sample mean of $\Re{\tr ( W_{\mathcal{A}} ) / N}$ gives an estimator for $\avg{\tr ( W_{\mathcal{A}} ) / N}$, and the variance of this estimator can be directly computed,
\begin{equation}
   \begin{split}
      &\text{Var}[\Re \tr ( W_{\mathcal{A}} ) / N] \\
      &\quad = \frac{1}{N^2} \left< \Re\tr ( W_{\mathcal{A}} )^2 \right> - \frac{1}{N^2} \Re \left< \tr ( W_{\mathcal{A}} ) \right>^2 \\
      &\quad = \frac{1}{2N^2} \left< \left| \tr ( W_{\mathcal{A}} )^2 \right| \right> + \frac{1}{2N^2} \left< \tr ( W_{\mathcal{A}} )^2 \right> \\
      &\qquad - \frac{1}{N^2} \Re \left< \tr ( W_{\mathcal{A}} ) \right>^2.
   \end{split}\label{eq:Wvar}
\end{equation}
The expectation values in the first and second terms in the variance can be factorized analogously to the Wilson loop expectation value, and are shown in Appendix~\ref{app:single-site-integrals} to be
\begin{equation} \label{eq:Wvar-terms}
\begin{aligned}
\left< \left| \tr ( W_{\mathcal{A}} )^2 \right| \right>
    &= 1 + (N^2-1) \langle \chi_{1,-1} \rangle^{A} \\
\left< \tr ( W_{\mathcal{A}} )^2 \right>
    &= \frac{N(N+1)}{2} \langle \chi_2 \rangle^A + \frac{N(N-1)}{2} \langle \chi_{1,1} \rangle^A,
\end{aligned}
\end{equation}
in terms of the single-site integrals $\avg{\chi_{1,-1}}$, $\avg{\chi_2}$, and $\avg{\chi_{1,1}}$, defined in Eqs.~\eqref{eq:w2def} and~\eqref{eq:w2pdef}. In total, the variance is
\begin{equation} \label{eq:Wvar-asymptotic}
\begin{split}
    \Var[\Re \tr ( W_{\mathcal{A}} ) / N] &=\frac{1}{2N^2} + \frac{O(e^{-c A})}{2N^2} - e^{-2 \sigma A},
\end{split}
\end{equation}
where $c$ is a constant.
The fact that $\left< \chi_{r} \right> < 1$ for non-trivial irreps $r$ (assuming that $g^2$ is finite)~\cite{Drouffe:1983fv} implies that $c>0$ and therefore that the variance is asymptotically constant as $A \rightarrow \infty$,
\begin{equation}
    \Var [\Re \tr ( W_{\mathcal{A}} ) / N] \sim \frac{1}{2N^2}.
\end{equation}

The signal-to-noise ratio for $n$ samples can be written exactly in terms of Eqs.~\eqref{eq:Wvar}, \eqref{eq:Wvar-terms}, and \eqref{eq:Wfac}, but for this analysis it is sufficient to identify the asymptotic behavior from Eqs.~\eqref{eq:Wvar-asymptotic} and \eqref{eq:Wfac}, giving
\begin{equation} \label{eq:SUNStN}
\begin{aligned}
    \text{StN}[\Re \tr ( W_{\mathcal{A}} ) / N] &= \frac{\avg{ \frac{1}{N} \tr \left( W_{\mathcal{A}} \right) }}{\sqrt{\frac{1}{n} \text{Var}[\Re \frac{1}{N} \tr (W_{\mathcal{A}})]}}\\
    &\sim N \sqrt{2 n} e^{-\sigma A},
\end{aligned}
\end{equation}
which degrades exponentially with area $A$. For the estimator $\Re \tr ( W_{\mathcal{A}} ) / N$, the analysis above shows that this can only be counteracted by exponentially increasing the number of samples $n$. Eqs.~\eqref{eq:Wvar-terms} and \eqref{eq:Wvar-asymptotic} also make clear that the leading asymptotic behavior of the variance is due to the typical magnitude-squared of the observable, $\left< |\tr(W_{\mathcal{A}})^2| / N^2 \right>$, which remains $O(1)$ for all areas. Cancellations due to phase fluctuations are required to reproduce the exponentially small Wilson loop expectation values predicted for large areas, confirming that the StN problem can be related to a sign problem in the Wilson loop observable.

Attributing the StN problem to $O(1)$ magnitudes for individual samples of the Wilson loop observable at all areas also inspires our deformations of the Wilson loop observable in the following sections. The quantity $\left< |\tr(W_{\mathcal{A}})^2|/N^2 \right>$ can be written as an integral of a non-holomorphic integrand which will generically be modified by contour deformations of the path integral. If we choose contour deformations that reduce the average magnitude of the observable, this quantity, and thus the leading term of the variance, will be reduced. The observable mean is unchanged and the StN ratio will thus increase under such a deformation.

For $SU(2)$, the single-site integrals can be evaluated straightforwardly (see Appendix~\ref{app:single-site-integrals}) and the $SU(2)$ variance is
\begin{equation}
   \begin{split}
      \text{Var} & [\Re \tr( W_{\mathcal{A}}^{SU(2)} ) / N] \\
      &= \frac{1}{4} + \frac{3}{4}\left( \frac{I_3(4/g^2)}{I_1(4/g^2)} \right)^A - e^{-2\sigma^{SU(2)} A},
   \end{split}\label{eq:SU2var}
\end{equation}
where $\sigma^{SU(2)}$ is given in Eq.~\eqref{eq:SU2sigma}.
The StN of $SU(2)$ Wilson loops in $(1+1)$D can therefore be explicitly calculated,
\begin{equation}
   \begin{split}
      \text{StN} & [\Re \tr(
      W_{\mathcal{A}}^{SU(2)} ) / N] \\&= \frac{ 2 \sqrt{n} e^{-\sigma^{SU(2)} A} }{\sqrt{ 1 + 3\left( \frac{I_3(4/g^2)}{I_1(4/g^2)} \right)^A  - 4 e^{-2\sigma^{SU(2)} A}}}.
   \end{split}\label{eq:SU2StN}
\end{equation}
Using numerical evaluation of the corresponding single-site integrals for $SU(N)$ Wilson loops yields theoretical curves for the variance and signal-to-noise for general $N$. In the studies below, we choose to deform the $(1,1)$ component of the Wilson loop, $W_{\mathcal{A}}^{11}$, instead of $\tr{W_{\mathcal{A}}}/N$ following the reasoning of Sec.~\ref{sec:rewriting}. The variance of $W_{\mathcal{A}}^{11}$ can be related to the variance of $\tr(W_{\mathcal{A}})/N$ and is compared to Monte Carlo results in the following sections.

\section{\texorpdfstring{$SU(2)$}{SU(2)} path integral contour deformations}\label{sec:su2}

As a proof-of-principle, we apply path integral contour deformations to Wilson loop calculations in $SU(2)$ lattice gauge theory in $(1+1)$D with open boundary conditions. An identical setting with gauge group $SU(3)$ is investigated in the following section.

\subsection{Gauge field parameterization} \label{sec:su2-param}

There are many possible parameterizations of the $SU(2)$ group manifold, any of which can be used to define valid path integral contour deformations.
We argue above that it is advantageous to consider a single component of the Wilson loop, taken without loss of generality to be $W_\mathcal{A}^{11}$, as the observable whose path integral contour is deformed in order to calculate $\left< W_{\mathcal{A}}^{11} \right> = \left< \tr(W_\mathcal{A})/N \right>$.
Contour deformations that reduce the magnitude of $W_\mathcal{A}^{11}$ in generic gauge field configurations while preserving $\left<W_\mathcal{A}^{11}\right>$ can be expected to reduce phase fluctuations and therefore the variance of $W_\mathcal{A}^{11}$.
The angular parameterization of each plaquette $P_x \in SU(2)$ is useful for this purpose, and is explicitly defined by
\begin{equation} \label{eq:su2-param}
\begin{aligned}
P_x^{11}&=\sin\theta_x\, e^{i\phi^1_x}, \\
P_x^{12}&=\cos\theta_x\, e^{i\phi^2_x}, \\
P_x^{21}&=-\cos\theta_x\, e^{-i \phi^2_x}, \\
P_x^{22}&=\sin\theta_x\, e^{-i\phi^1_x},
\end{aligned}
\end{equation}
following the generalized $SU(N)$ angular parameterization given in Ref.~\cite{Bronzan:1988wa}. The azimuthal angles satisfy $\phi^1_x,\phi^2_x \in [0, 2\pi]$, with endpoints identified, while the angle $\theta_x$ spans the finite interval $[0,\pi/2]$. The normalized Haar measure can be written in these coordinates as
\begin{equation} \label{eq:su2-haar}
\begin{aligned}
dP_x &= h(\Omega_x) \ d\Omega_x
\equiv \frac{1}{4 \pi^2} \sin(2 \theta_x) \ d\theta_x \ d\phi^1_x \ d\phi^2_x.
\end{aligned}
\end{equation}

We begin by considering the effects of simple deformations using these parameters. In the simplest case of a region $\mathcal{A}$ with area $A=1$, the Wilson loop consists of a single plaquette, $W_{\mathcal{A}}^{11} = P_x^{11}$, where the loop starts and ends at site $x$. The magnitude of $W_\mathcal{A}^{11}$ can be reduced by $e^{-\lambda}$ by deforming $\phi^1_x \rightarrow \phi^1_x + i\lambda$ analogously to the approach described for $e^{i\phi}$ integrals above.
In the case of $A=2$, the Wilson loop can be written in terms of the product of two plaquettes, $W_\mathcal{A}^{11} = (P_x P_{x'})^{11}$. In the angular parameterization, the Wilson loop is a sum of two terms
\begin{equation}
    (P_{x} P_{x'})^{11} = \sin\theta_x \sin\theta_{x'} e^{i \phi_x^1 + i\phi^{1}_{x'}} + \cos\theta_x \cos\theta_{x'} e^{i\phi_x^2 - i\phi^{2}_{x'}}.
\end{equation}
The first term involves products of diagonal entries whose magnitude can be reduced by $e^{-\lambda}$ by taking $\phi_x^1 \rightarrow \phi_x^1 + i \lambda $ or $\phi^{1}_{x'} \rightarrow \phi^{1}_{x'} + i \lambda$ and the second term involves off-diagonal components whose magnitude can be reduced analogously by taking $\phi_x^2 - \phi^{2}_{x'} \rightarrow (\phi_x^2 - \phi^{2}_{x'}) + i\lambda$. 
For $A>2$, it can be seen similarly that shifting $\phi^1_x \rightarrow  \phi^1_x + i \lambda$ and $(\phi^2_{x} - \phi^2_{x+1}) \rightarrow (\phi^2_{x} - \phi^2_{x+1})  + i \lambda$ for all $x$ leads to suppression of the magnitudes of all terms appearing in $W_\mathcal{A}^{11}$.

A family of contour deformations capable of expressing these constant imaginary shifts to the phases of all elements of $P_{x}$ can therefore be expected to reduce phase fluctuations and the variance of $W_\mathcal{A}^{11}$.
Such a family of contour deformations is parameterized below as a subset of the vertical deformation expanded in a Fourier series in Eqs.~\eqref{eq:vertical-deform}--\eqref{eq:vertical-fourier-deform2}.

An alternative parameterization of $SU(2)$ is explored in Appendix~\ref{app:su2-euler}, in which it is found that imaginary shifts along these lines are more difficult to express and orders of magnitude less variance reduction is achieved when applying the same optimization methods. This exploration suggests that a choice of parameterization that allows the observable to be expressed in the form $e^{i\phi}$ is important for variance reduction in observables afflicted with a sign problem.

It is also possible to directly parameterize the gauge field $U_{x,\mu} \in SU(2)$ using Eq.~\eqref{eq:su2-param}.
This alternative parameterization may be useful in more than two spacetime dimensions, where Gauss' Law constraints imply that not all plaquettes are independent and a path integral change of variables from $U_{x,\mu}$ to $P^{\mu\nu}_x$ cannot be performed straightforwardly.

\subsection{Fourier deformation basis} \label{sec:su2-fourier}

In our study of $SU(2)$ gauge theory, we optimize over a family of vertical contour deformations expressed in terms of a Fourier series truncated above a specific cutoff mode. To avoid a costly Jacobian calculation, each plaquette variable $P_x$ is deformed conditioned on plaquettes $P_y$ at sites earlier in the product defining $W_{\mathcal{A}}$ in Eq.~\eqref{eq:Wfac1}, which we write as $y \leq x$.
This family of deformations is given by
\begin{equation}
    \begin{split}
        \tilde{\theta}_{x} &\equiv \theta_{x} + i \sum_{y\leq x} f_\theta(\theta_{y},\phi^1_{y},\phi^2_{y};\kappa^{xy}, \lambda^{xy}, \eta^{xy}, \chi^{xy}, \zeta^{xy}),    \\
        \tilde{\phi}^1_{x} &\equiv  \phi^1_{x} + i \kappa_0^{x;\phi^1}  \\
        &\hspace{20pt} + i \sum_{y\leq x} f_{\phi^1}(\theta_{y},\phi^1_{y},\phi^2_{y};\kappa^{xy}, \lambda^{xy}, \eta^{xy}, \chi^{xy}, \zeta^{xy}), \\
        \tilde{\phi}^2_{x} &\equiv  \phi^2_{x} + i \kappa_0^{x;\phi^2} \\
        &\hspace{20pt}+ i \sum_{y\leq x} f_{\phi^2}(\theta_{y},\phi^1_{y},\phi^2_{y};\kappa^{xy}, \lambda^{xy}, \eta^{xy}, \chi^{xy}, \zeta^{xy}),
    \end{split}\label{eq:uppertridef}
\end{equation}
in terms of parameters $\kappa^{xy}$, $\lambda^{xy}$, $\eta^{xy}$, $\chi^{xy}$, and $\zeta^{xy}$. The functions $f_\theta$, $f_{\phi^1}$, and $f_{\phi^2}$ compute the shift in the imaginary direction of the angular parameters of $P_x$ conditioned on $P_y$, and their decomposition in terms of Fourier modes is detailed below. For this ordered dependence on previous sites, the Jacobian determinant factorizes into a product of block determinants per lattice site
\begin{equation}
    \begin{split}
        J &= \prod_{x} j_{x}(\theta_{x},\phi^1_{x},\phi^2_{x}),
    \end{split}\label{eq:Jupper}
\end{equation}
where
\begin{equation}
  j_{x}(\theta_{x},\phi^1_{x},\phi^2_{x}) = \text{det}\begin{pmatrix} \frac{\partial f_\theta}{\partial \theta_{x}} & \frac{\partial  f_\theta}{\partial \phi^1_{x}} & \frac{\partial f_\theta}{\partial \phi^2_{x}} \\
     \frac{\partial f_{\phi^1}}{\partial \theta_{x}}   &  \frac{ \partial f_{\phi^1} }{\partial \phi^1_{x}} &  \frac{\partial f_{\phi^1}}{\partial \phi^2_{x}} \\
        \frac{\partial f_{\phi^2}}{\partial \theta_{x}}   &  \frac{\partial f_{\phi^2}}{\partial \phi^1_{x}} &  \frac{\partial f_{\phi^2}}{\partial \phi^2_{x}} \end{pmatrix}. \label{eq:jupper}
\end{equation}
The structure of the deformation in Eq.~\eqref{eq:uppertridef} therefore bypasses the need for expensive Jacobian determinant calculations involving matrices whose rank grows with the spacetime volume and is inspired by analogous methods to simplify Jacobian determinant calculations in normalizing flows~\cite{papamakarios2019normalizing}. Note that an absolute value is not taken over the determinant in Eq.~\eqref{eq:jupper}.

The vertical deformation in Eq.~\eqref{eq:uppertridef} can be expanded in a multi-parameter Fourier series as
\begin{widetext}
\begin{equation}
    \begin{split}
        f_\theta &= \sum_{m=1}^{\Lambda}  \kappa_m^{xy;\theta} \sin\left(2 m \theta_{y} \right) \left\lbrace 1  + \sum_{n = 1}^{\Lambda} \left[ \eta_{mn}^{xy;\theta, \phi^1}  \sin(n \phi^1_{y } + \chi_{mn}^{xy;\theta,\phi^1}) + \eta_{mn}^{xy;\theta,\phi^2} \sin(n\phi^2_{y} + \chi_{mn}^{xy;\theta,\phi^2}) \right] \right\rbrace,   \\
        f_{\phi^1} &= \sum_{m=1}^{\Lambda}  \kappa_{m}^{xy;\phi^1} \sin (m \phi^1_{y} + \zeta_{m}^{xy;\phi^1} )      \left\lbrace 1 + \sum_{n=1}^{\Lambda} \left[ \lambda_{mn}^{xy;\phi^1,\theta}  \sin(2 n \theta_{y} ) + \eta_{mn}^{xy;\phi^1,\phi^2}  \sin(n\phi^2_{y}    + \chi_{mn}^{xy;\phi^1,\phi^2}) \right]  \right\rbrace, \\
        f_{\phi^2} &= \sum_{m=1}^{\Lambda} \kappa_m^{xy;\phi^2} \sin(m\phi^2_{y} + \zeta_m^{xy;\phi^2}) \left\lbrace 1 + \sum_{n=1}^{\Lambda} \left[ \lambda_{mn}^{xy;\phi^2,\theta} \sin(2 n \theta_{y}) + \eta_{mn}^{xy;\phi^2, \phi^1} \sin(n \phi^1_{y}    + \chi_{mn}^{xy;\phi^2,\phi^1})   \right] \right\rbrace,
    \end{split}
    \label{eq:nonlocaltransform}
\end{equation}
\end{widetext}
where $\Lambda$ is a hyperparameter that sets the maximum Fourier mode to include and controls the total number of free parameters. As the zero modes have trivial $y$ dependence, we have collected them in Eq.~\eqref{eq:uppertridef} into the $y$-independent terms $\kappa_0^{x; \phi^{1}}$ and $\kappa_0^{x; \phi^{2}}$. The included Fourier terms are defined to satisfy the constraints $\widetilde{\theta}_{x}(0) = 0$, $\widetilde{\theta}_{x}(\pi/2) = \pi/2$,  $\widetilde{\phi}^1_{x}(0) = \widetilde{\phi}^1_{x}(2\pi)$, and $\widetilde{\phi}^2_{x}(0) = \widetilde{\phi}^2_{x}(2\pi)$, which together ensure that the endpoints of both the zenith and azimuthal integration domains are appropriately handled as described in Sec.~\ref{sec:SUNdef}.
The derivatives needed for the Jacobian in Eqs.~\eqref{eq:Jupper}--\eqref{eq:jupper} can be calculated straightforwardly by differentiating Eq.~\eqref{eq:nonlocaltransform}.
The additional factor describing the change in the Haar measure needed to compute the Jacobian of the group measure is given in these coordinates as
\begin{equation} \label{eq:su2-haar-ratio}
    \prod_{x} \frac{h(\widetilde{\Omega}_x)}{h(\Omega_x)} = \prod_{x} \left[ \frac{\sin(2\tilde{\theta}_{x})}{\sin(2\theta_{x})} \right] .
\end{equation}

Combining the results of Eq.~\eqref{eq:su2-param} and Eqs.~\eqref{eq:uppertridef}--\eqref{eq:su2-haar-ratio}, the expectation value of any holomorphic observable $\obs(\{ P_x \})$ is equal to the expectation value of the deformed observable
\begin{equation} \label{eq:su2-deformed-obs}
    Q(\{ P_x \}) \equiv \obs(\{ \widetilde{P}_x \}) \frac{e^{-S(\{ \widetilde{P}_x \})}}{e^{-S(\{ P_x \})}} \prod_x j_x \left[ \frac{\sin(2\tilde{\theta}_{x})}{\sin(2\theta_{x})} \right],
\end{equation}
where
\begin{equation}
    \widetilde{P}_x = \left(\begin{matrix}
    \sin\widetilde{\theta}_x e^{i \widetilde{\phi}^1_x} & 
    \cos\widetilde{\theta}_x e^{i \widetilde{\phi}^2_x} \\
    -\cos\widetilde{\theta}_x e^{-i \widetilde{\phi}^2_x} &
    \sin\widetilde{\theta}_x e^{-i \widetilde{\phi}^1_x}
    \end{matrix}\right) \in SL(2,\mathbb{C}).
\end{equation}
If the plaquettes are sampled in the matrix representation for Monte Carlo evaluation, computing the observable $Q$ in Eq.~\eqref{eq:su2-deformed-obs} requires converting to the angular representation before deforming and evaluating. This conversion is given by
\begin{equation} \label{eq:su2-param-bwd}
\begin{aligned}
    \theta_x &= \arcsin(|P^{11}_x|), \\
    \phi^1_x &= \arg(P^{11}_x), \\
    \phi^2_x &= \arg(P^{12}_x),
\end{aligned}
\end{equation}
and can be done when evaluating the observable $Q(\{ P_x \})$. Though these functions are not entire, the conversion used here does not determine whether the integrand itself is a holomorphic function of these angular parameters.

\subsection{Optimization procedure} \label{sec:su2-training}
This contour deformation expanded in a Fourier series provides a means of exploring deformed observables with potentially reduced variance. It is shown above that simple deformations within this family are possible to construct by hand and are already sufficient to reduce the average magnitude of Wilson loop observables. However, these deformations are restricted to zero modes of the Fourier expansion and rely on construction based on intuition. To maximize the variance reduction, we explore numerical optimization of the deformation parameters $\kappa^{xy}$, $\lambda^{xy}$, $\eta^{xy}$, $\chi^{xy}$, and $\zeta^{xy}$ as discussed in Sec.~\ref{sec:obs-deforms}. We are interested in $\Re W_{\mathcal{A}}^{11}$, for which the terms of Eq.~\eqref{eq:Wvar} that can be modified by contour deformation are
\begin{equation} \label{eq:loss-fn}
    \mathcal{L} \equiv \left< (\Re Q_{\mathcal{A}})^2 \right> = \frac{1}{2}\left< |Q_{\mathcal{A}}^2| \right> + \frac{1}{2}\left< Q_{\mathcal{A}}^2 \right>,
\end{equation}
where $Q_{\mathcal{A}}$ is the deformed observable associated with the $W_{\mathcal{A}}^{11}$. The first term in Eq.~\eqref{eq:loss-fn} is manifestly non-holomorphic due to the absolute value over a complex-valued observable, while the second term includes squared reweighting factors of the original and deformed action which prevent identification as a deformation of $\left<(W_{\mathcal{A}}^{11})^2 \right>$. These terms together define the \emph{loss function} $\mathcal{L}$ that we aim to minimize as a function of the deformation parameters.

This loss function is written as an expectation value in terms of sampling from the original Monte Carlo weights $e^{-S(\{P_x\})}$, and its gradient can similarly be written as an expectation value,
\begin{equation} \label{eq:loss-grad}
    \nabla \mathcal{L}
    = \left< 2 \Re Q_{\mathcal{A}} \nabla \Re Q_{\mathcal{A}} \right>.
\end{equation}
The term $\nabla \Re Q_{\mathcal{A}}$ can be expanded using the explicit form of $Q_{\mathcal{A}}$ given in Eq.~\eqref{eq:su2-deformed-obs}, as well as the forms of the observable $W_{\mathcal{A}}^{11}$ and the action $S$ in terms of $\{P_x\}$ in Sec.~\ref{sec:suN-defns}. For this study, the gradient $\nabla \Re Q_{\mathcal{A}}$ was computed explicitly and cross-checked using automatic differentiation available in the JAX library~\cite{jax2018github}. Eq.~\eqref{eq:loss-grad} can be stochastically estimated using an (undeformed) ensemble of $n$ configurations $\{ P^{i}_x \}$, $i \in [1, \dots, n]$, drawn proportional to the weight $e^{-S(\{P_x\})}$,
\begin{equation} \label{eq:sample-loss-grad}
    \nabla \mathcal{L} \approx \frac{1}{n} \sum_{i=1}^{n} \left[ 2 \Re Q_{\mathcal{A}}(\{P^i_x\}) \nabla \Re Q_{\mathcal{A}}(\{P^i_x\}) \right].
\end{equation}

In all experiments below, we used the Adam optimizer~\cite{kingma2017adam} to iteratively update parameters based on these stochastic gradient estimates. Each gradient estimate was computed using $1/100$th of the generated ensemble; empirically, this small subset of the data was sufficient to learn useful manifold parameters with significant variance reduction. The optimizer was configured with default hyperparameters, except for a dynamically scheduled step size. Stochastic noise on gradient estimates and large optimizer step size can either slow convergence or result in unstable training dynamics, while step sizes that are too small waste computation as parameters fail to move quickly along precisely estimated gradients. We thus used a dynamic schedule that reduced the step size over time. In particular, our step size schedule started with an initial step size $s_0$ and then permanently reduced the step size by a factor of $F$ (i.e.~$s_{i+1} = s_i / F$) each time the loss function failed to improve over a window of $W$ steps. The schedule halted training after the step size was reduced $N_r$ times. We used the parameters $F = 10$, $W = 50$, $N_r = 2$, and $s_0 = 10^{-2}$ for both $SU(2)$ and $SU(3)$ gauge theory.

\begin{figure*}
    \centering
    \includegraphics{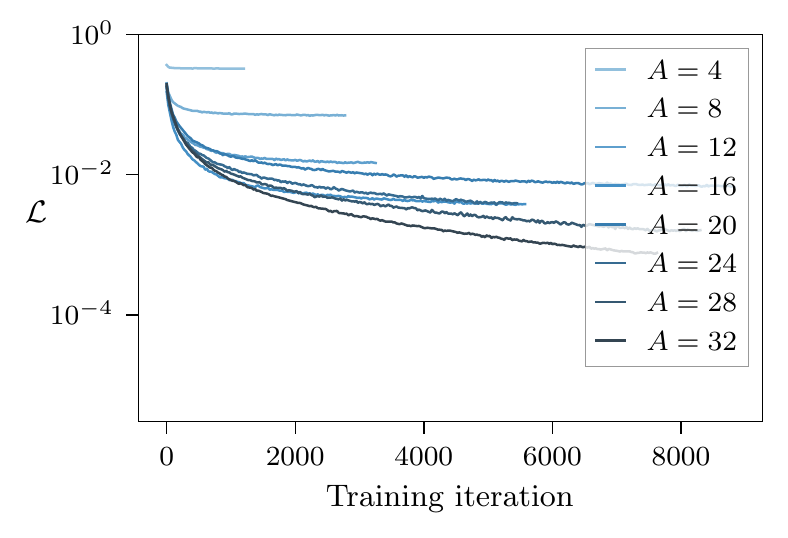}
    \includegraphics{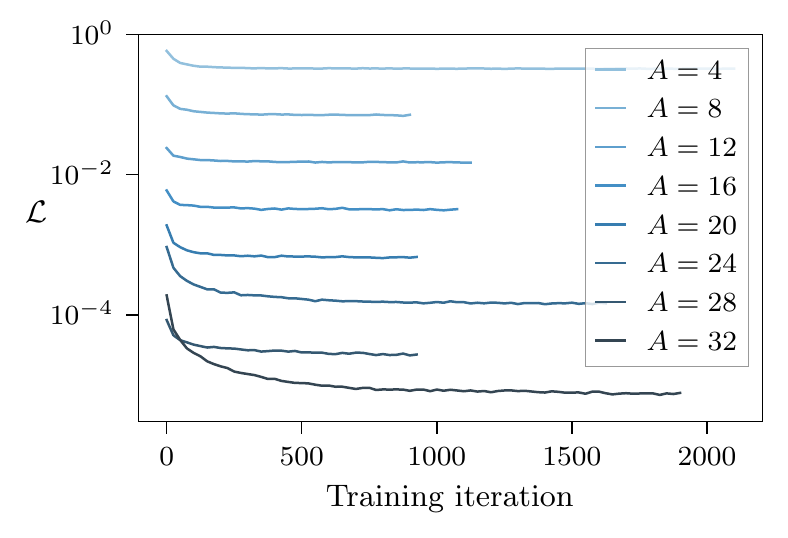}
    \caption{$SU(2)$ Wilson loop loss $\mathcal{L}$ plotted with respect to training iterations for optimization of Wilson loops with $g=0.71$ and area $A$ starting with the undeformed manifold (left) or starting with the deformed manifold calculated for area $A-1$ (right). The loss curves are averaged over blocks of 25 iterations for clarity. In each case, the training is halted based on the plateau criterion described in the main text, resulting in traces of different lengths. For Wilson loops with larger areas, the final loss value is substantially lower when initialized from optimal manifold parameters at a $A-1$, despite using nearly four times fewer training iterations.}
    \label{fig:su2_Euler_training}
\end{figure*}

In preliminary investigations, we found that naive manifold optimization resulted in overtraining, i.e.~overfitting parameters to the specific training data available~\cite{hardt2016train,pmlr-v48-lina16,zhang2018study}. In the context of manifold optimization, this corresponds to minimizing a finite-ensemble variance estimator rather than minimizing the true variance of $\Re Q_{\mathcal{A}}$. In practice this produced deformed observables with higher variance when measured on a different ensemble than the training set.

To mitigate overtraining in the final results, we reserved a ``test set'' of configurations, independent of the training data, on which the loss function was periodically measured~\cite{raschka2020model}; the reserved test set of configurations was chosen to have the same size as the training set. The step size schedule was configured to use loss measurements based on this test set, ensuring that training was slowed and halted before significantly overfitting. We further used a mini-batching technique, in which a set of $m \leq n$ configurations are resampled from the training set to estimate Eq.~\eqref{eq:sample-loss-grad}, as a means of avoiding overtraining~\cite{bottou2018optimization}. The mini-batch size was chosen to be equal to the size of the full training set (i.e.~$m = n$), thus mini-batch evaluation in this context was just a resampling operation, giving variation in gradient estimates over multiple evaluations. The fluctuations in these gradient estimates are on the order of the uncertainty of the variance estimator, preventing overfitting below this uncertainty. For each observable $W_{\mathcal{A}}^{11}$ we also found it important to restrict to deforming only the plaquettes within the region $\mathcal{A}$. Though including extra degrees of freedom cannot make the optimal variance any higher (the optimization steps could always leave those plaquettes undeformed), in practice we found that including such degrees of freedom allowed the deformed manifold to rapidly overtrain, resulting in worse performance overall. Appendix~\ref{sec:loss-reg} details further possible approaches to avoiding overfitting and overlap problems using a regularizing term added to the loss function. These approaches were found to be unnecessary for our final results.

Finally, making a good choice of initial manifold parameters yielded practical improvement in training time and quality. On one hand, initializing the manifold parameters to zero ensures that gradient descent starts from the original manifold, and the optimization procedure should find a local minimum with variance no higher than the original manifold (up to noise from stochastic gradient estimates). However, correlations in sign and magnitude fluctuations of observables with similar structure, such as Wilson loops $W_{\mathcal{A}}^{11}$ and $W_{\mathcal{A}'}^{11}$ with overlapping areas $\mathcal{A}$ and $\mathcal{A}'$, suggest that the variance reduction from contour deformations will be correlated as well. Though the optimal manifold for one observable will not generically be optimal for the other, it can serve as a better starting point than the original manifold. In our study of Wilson loops, we initialized manifold parameters for each Wilson loop of area $A$ using the optimal parameters for the Wilson loop of area $A-1$, when the Fourier cutoff $\Lambda = 0$, or using the optimal parameters for the Wilson loop of area $A$ and cutoff $\Lambda-1$, when $\Lambda \neq 0$. Figure~\ref{fig:su2_Euler_training} shows the improvement in optimization time and quality using this method for manifold deformations with $\Lambda = 0$. While this approach sacrifices the guarantee that the local minimum obtained corresponds to a deformed observable with variance no higher than the original observable (in the limit of infinitely precise gradient estimates), in practice we find that this property is not violated. We also note that this property can be easily checked after optimization, and if the variance were found to increase with respect to the original observable training could instead be started from the original manifold to recover the guarantee.

\subsection{Monte Carlo calculations} \label{sec:su2-mc}

We investigated the practical performance of these contour deformations on the three sets of $SU(2)$ parameters detailed in Table~\ref{tab:couplings}. These parameters correspond to variation in the string tension by a factor of 4. At each choice of parameters, $n = 32000$ configurations were generated, exploiting the factorization of the path integral to draw samples of each plaquette $P_x$ independently from the marginal distribution,
\begin{equation}
    p(P_x) \propto e^{-\frac{1}{g^2} \tr (P_x + P_x^{-1})}.
\end{equation}
Plaquette samples were generated using Hybrid Monte Carlo~\cite{Duane:1987de} with parameters tuned to maintain autocorrelation times of order 1--2,\footnote{
For $SU(2)$ gauge theory it is also possible to apply a heat bath algorithm to directly draw samples. However, more complicated variants are required for $SU(3)$~\cite{Creutz:1980zw,Pietarinen:1981cd,deForcrand:2005xr} and Hybrid Monte Carlo was selected for simplicity.} and these individual plaquettes were then arranged into lattices consisting of $V$ sites each. A random shuffle was applied to the collection of plaquettes prior to this assignment to avoid spurious spatial correlations, ensuring an asymptotically vanishing bias in the limit of infinite ensemble size.

\begin{figure*}
    \centering
    \includegraphics{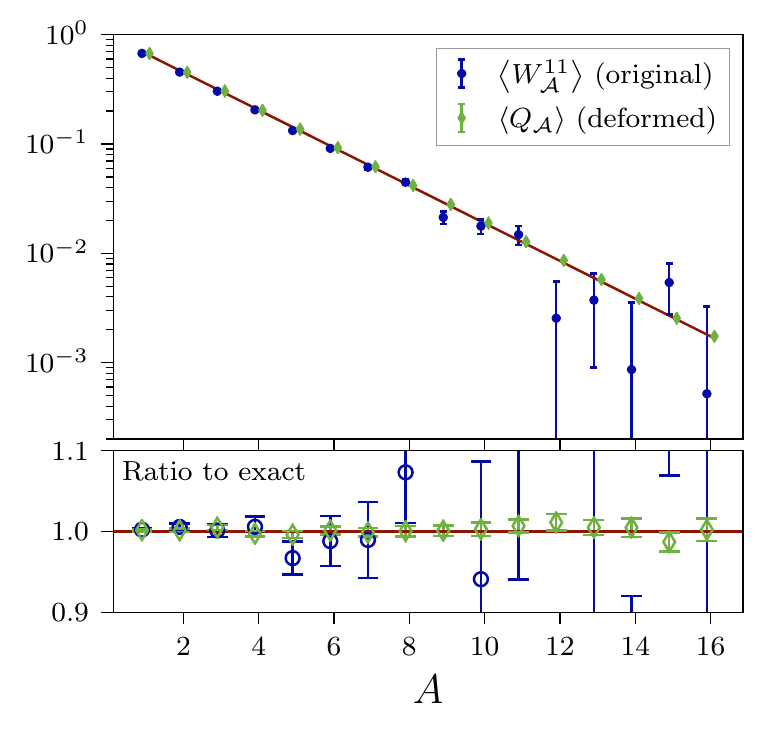}
    \includegraphics{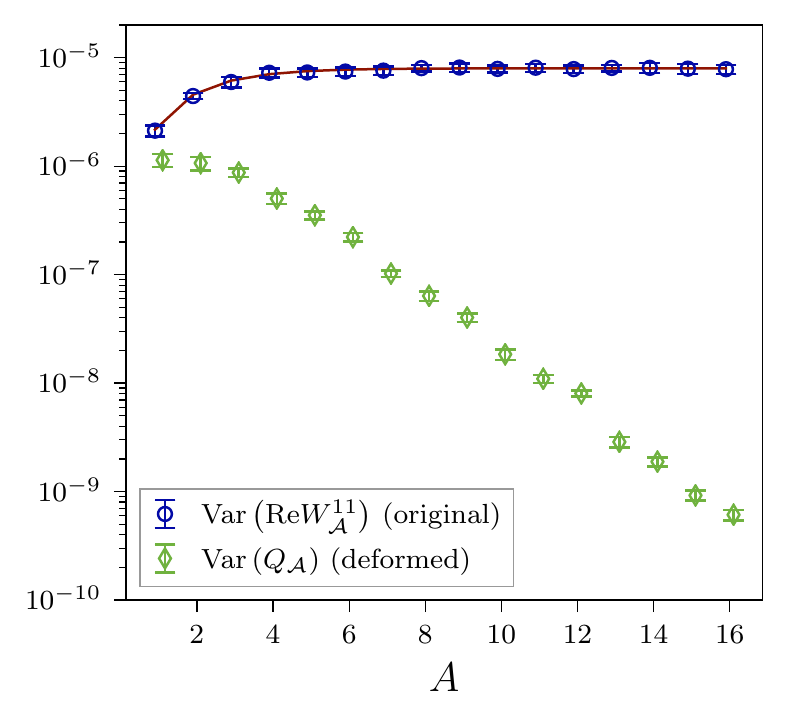}
    \includegraphics{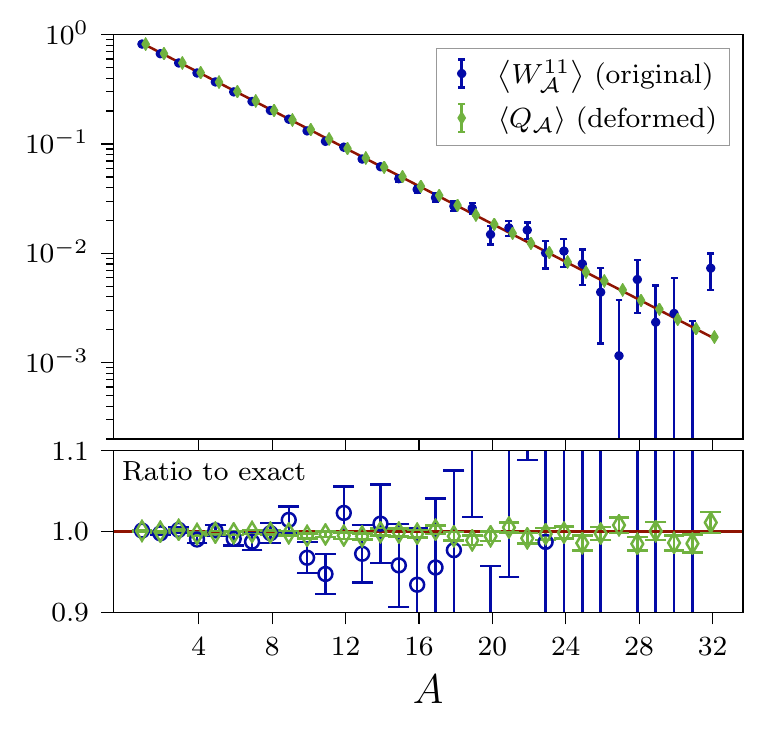}
    \includegraphics{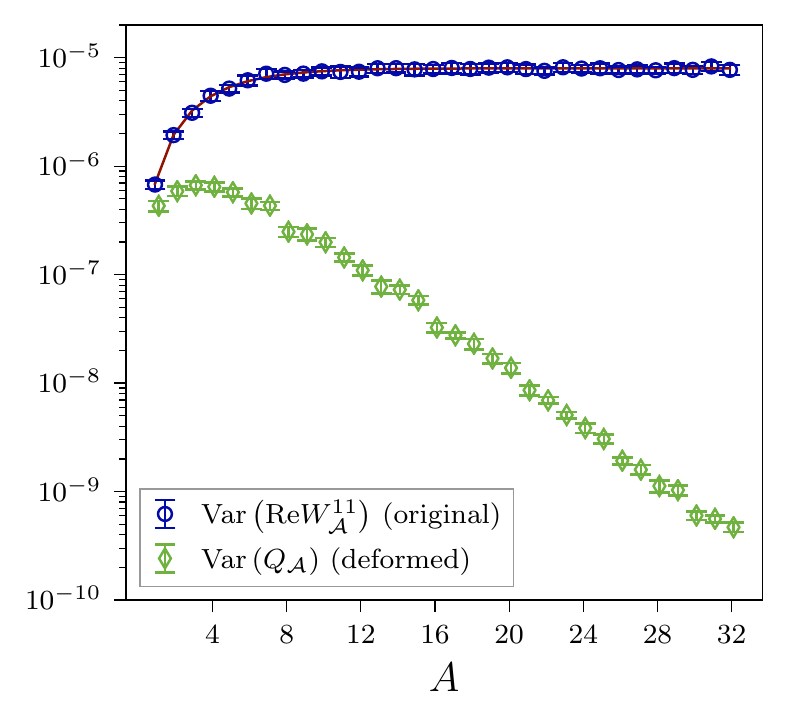}
    \includegraphics{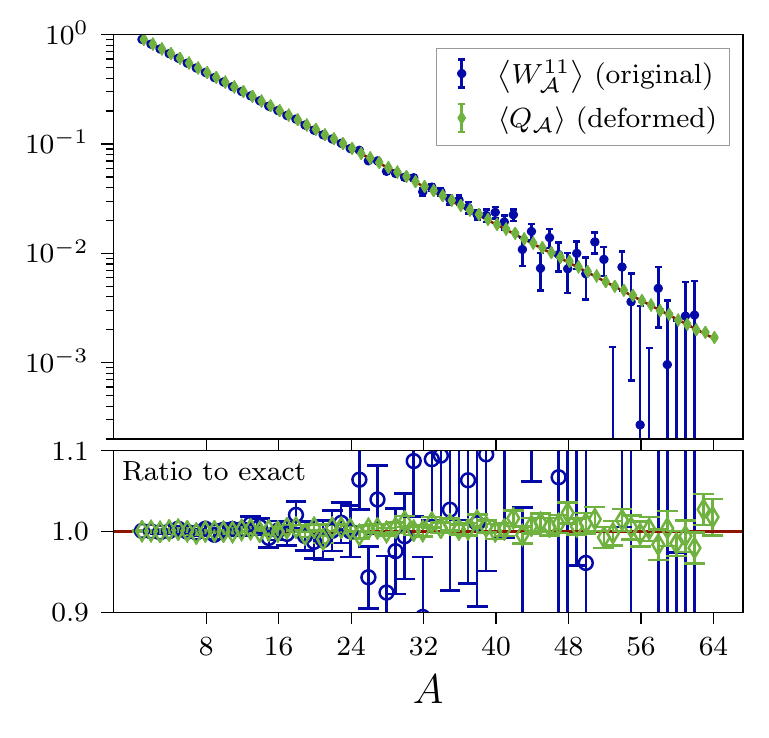}
    \includegraphics{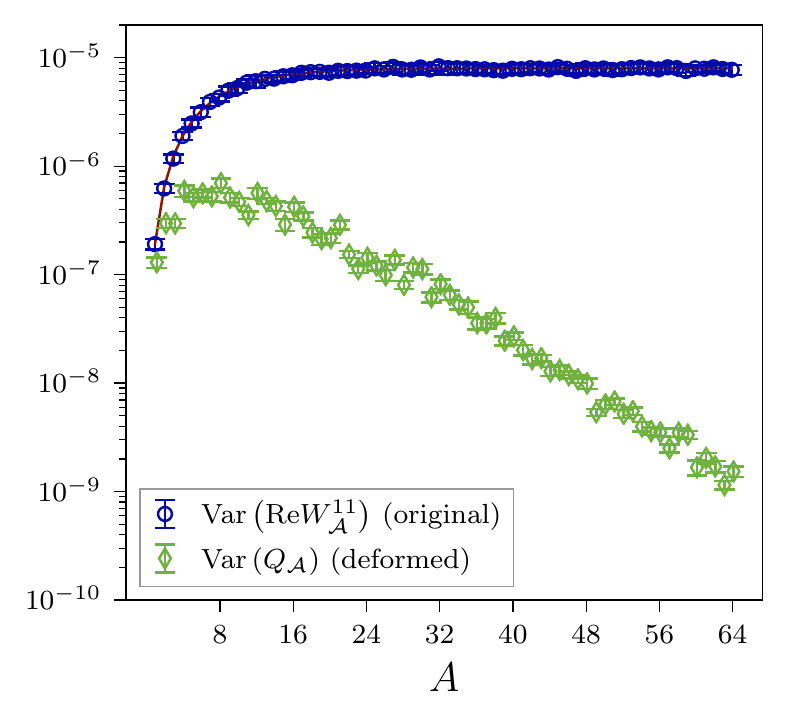}
    \caption{$SU(2)$ Wilson loop expectation values and variances for ensembles with three different values of the gauge coupling $g=0.98,\ 0.71,\ 0.51$ (top to bottom). Red lines indicate analytical results for $\left< W_{\mathcal{A}}^{11} \right> = \left< \tr(W_{\mathcal{A}})/2 \right>$ (left plots) and for $\Var(\Re W_{\mathcal{A}}^{11})$ (right plots).}
    \label{fig:su2_Euler_observable}
\end{figure*}

On each ensemble, we performed a study of deformations with Fourier cutoff $\Lambda$ fixed to zero. For all three choices of parameters, training on a subset of $320$ configurations was sufficient to converge to manifolds with variance reduction up to four orders of magnitude when comparing the largest deformed observables to the original Wilson loop observables. An additional subset of $320$ configurations was reserved as the test set during optimization, and the remaining 31360 configurations were used to measure results. Measurements of $Q_{\mathcal{A}}$ were found to be consistent with the exact results given in Sec.~\ref{sec:suN-defns} and with the Monte Carlo results for $W_{\mathcal{A}}^{11}$ in the region where the estimates of $W_{\mathcal{A}}^{11}$ were reliable. The variance of $\Re W_{\mathcal{A}}^{11}$ is dominated by $\left< (\Re W_{\mathcal{A}}^{11})^2 \right>$, which is positive-definite for all gauge field configurations. It was therefore possible to precisely measure the variance without a sign/StN problem, and the expected agreement with analytical results was reproduced. In particular, $O(1)$ scaling at large area can be clearly seen. In contrast, we found that the variance of $Q_{\mathcal{A}}$ for manifolds optimized as above falls exponentially at large area, giving exponential improvements in the signal-to-noise. These results are presented for all three ensembles in Fig.~\ref{fig:su2_Euler_observable}.

\begin{figure}
    \centering
    \includegraphics{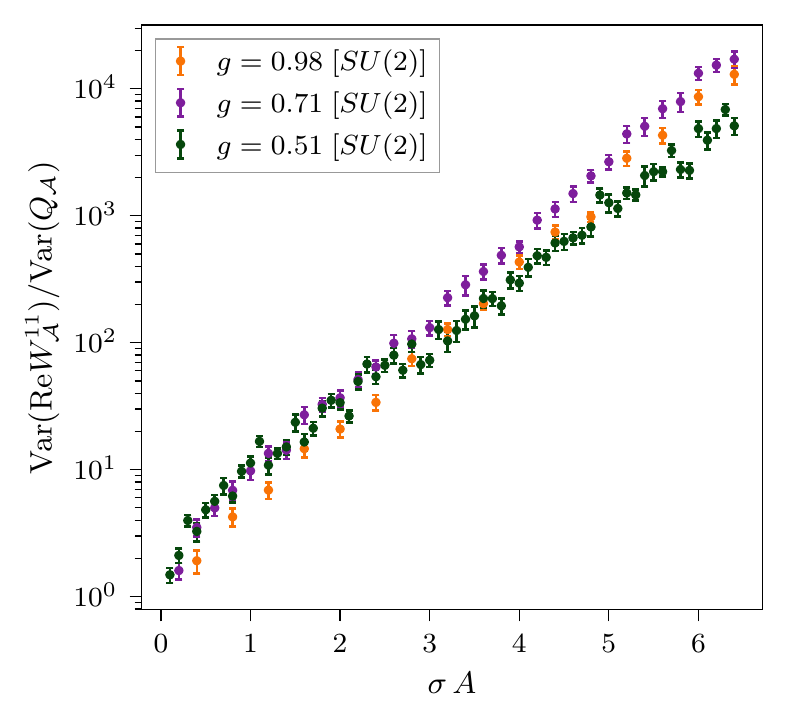}
    \caption{$SU(2)$ Wilson loop variance ratios of standard observables to deformed observables for ensembles with three different values of the gauge coupling $g$, corresponding to three different values of lattice spacing.}
    \label{fig:su2_Euler_lattice}
\end{figure}

Improvements from contour deformation were also found to be similar for Wilson loops with fixed physical area $\sigma A$ across the three values of the lattice spacing. Fig.~\ref{fig:su2_Euler_lattice} compares the improvement in the Wilson loop variance versus the dimensionless scale $\sigma A$ across all three ensembles, and the three curves can be seen to nearly collapse at small areas, though there are differences of roughly a factor of 4 at the largest areas. Despite this variation, the variance of the Wilson loop with largest area is reduced by more than $10^3$ even for the finest ensemble. Analogous results were observed for $(1+1)$D $U(1)$ gauge theory in Ref.~\cite{Detmold:2020ncp}.

\begin{figure}
    \centering
    \includegraphics{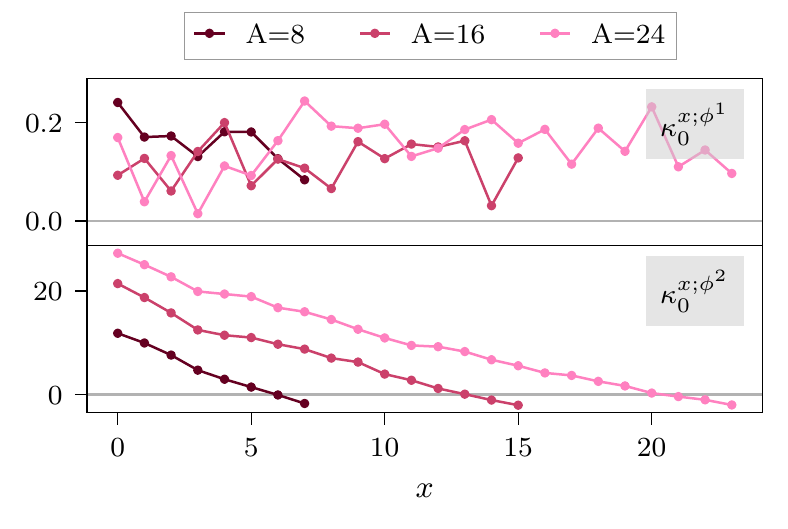}
    \caption{The manifold parameters found by optimizing the variance of the deformed Wilson loop observable $Q_{\mathcal{A}}$ at three different choices of area $A$ on the ensemble with total volume $V=32$ and $\beta = 8.0$. Optimization at each $A$ was initialized using the parameters found for the observable with area $A-1$, as detailed in the main text.}
    \label{fig:su2_Euler_manifold}
\end{figure}

For $\Lambda = 0$, there are few enough parameters that it is possible to investigate the optimal parameters found by the optimization procedure. Fig.~\ref{fig:su2_Euler_manifold} depicts the values of $\kappa_0^{x;\phi^{1}}$ and $\kappa_0^{x;\phi^{2}}$ when optimized to reduce variance of $Q_{\mathcal{A}}$ at three choices of the loop area. It is argued in Sec.~\ref{sec:su2-param} that the magnitude of $Q_{\mathcal{A}}$ can be reduced on each sample if $\phi^{1}_x$ and the differences $\phi^{2}_{x} - \phi^{2}_{x+1}$ are shifted by a positive imaginary constant. This manifold is approximately discovered by the optimization procedure: the final parameters are a positive, nearly constant $\kappa_0^{x;\phi^{1}}$, corresponding to a positive imaginary shift applied to $\phi^{1}_x$, and a decreasing $\kappa_0^{x;\phi^{2}}$, corresponding to a positive imaginary shift applied to each difference $\phi^{2}_x - \phi^{2}_{x+1}$. Only these relative differences between neighboring $\phi^{2}_x$ have an effect on the value of $Q_{\mathcal{A}}$, thus the overall shift on the collection of $\kappa_0^{x;\phi^{2}}$ is irrelevant.

\begin{figure}
    \centering
    \includegraphics{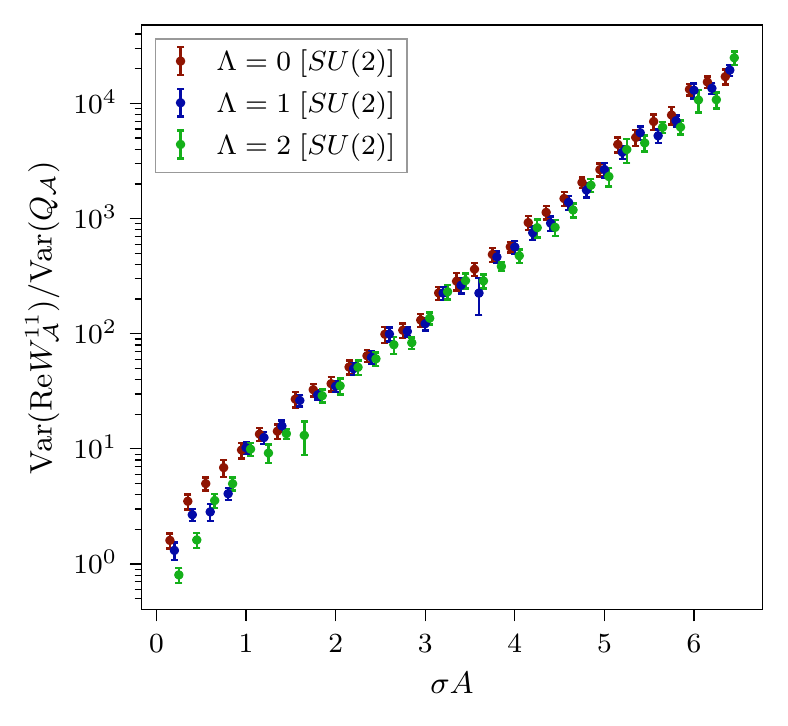}
    \caption{$SU(2)$ Wilson loop variance ratios of standard observables to deformed observables for ensembles with $g=0.71$ and three different values of the manifold parameterization cutoff.}
    \label{fig:su2_Euler_cutoff}
\end{figure}

We further optimized manifold parameters using Fourier cutoffs $\Lambda = 1, 2$ on the ensemble with coupling $g=0.71$ to investigate gains from including higher Fourier modes. Including Fourier modes larger than the constant term enables more complex deformations of each angular parameter, and introduces possible dependence on plaquettes at sites $y \leq x$ when deforming $P_x$. Despite this increased expressivity, these additional parameters did not provide significantly larger StN improvements compared to using only constant terms, as shown in Fig.~\ref{fig:su2_Euler_cutoff}. In some cases, the optimized manifold with larger cutoff resulted in higher variance (lower variance ratio) than the optimized manifold with cutoff $\Lambda = 0$. The manifolds with larger cutoffs include all possible manifolds with smaller cutoffs, thus this is necessarily a training effect, likely due to noisier gradients and less stable training dynamics. We did not pursue noise reduction and alternative approaches to training (such as iteratively including higher $\Lambda$) as these manifolds with higher cutoffs did not produce significant improvements at any value of the area.

\section{\texorpdfstring{$SU(3)$}{SU(3)} path integral contour deformations}\label{sec:su3}

We further investigated the ability of contour deformations to reduce the variance of Wilson loops in $SU(3)$ gauge theory in $(1+1)$D with open boundary conditions. This setting is identical to the previous section, except for the use of $SU(3)$ rather than $SU(2)$ gauge field variables. Suitable parameterizations for contour deformations of $SU(3)$ gauge fields are discussed below.

\subsection{Gauge field parameterization and contour deformation} \label{sec:su3-param}

For the $SU(3)$ gauge group, we use the angular parameterization constructed in Ref.~\cite{Bronzan:1988wa}. The components of a single plaquette $P_x \in SU(3)$ are parameterized as
\begin{equation}
    \begin{split}
        P_x^{11}&=\cos\theta_x^1\cos\theta_x^2 e^{i\phi_x^1}, \\ P_x^{12}&=\sin\theta_x^1 e^{i\phi_x^3}, \\ P_x^{13}&=\cos\theta_x^1\sin\theta_x^2 e^{i\phi_x^4}\, ,\\
P_x^{21}&=\sin\theta_x^2\sin\theta_x^3 e^{-i(\phi_x^4+\phi_x^5)} \\
&\hspace{20pt} -\sin\theta_x^1\cos\theta_x^2\cos\theta_x^3 e^{i(\phi_x^1+\phi_x^2-\phi_x^3)}\, ,\\
P_x^{22}&=\cos\theta_x^1\cos\theta_x^3 e^{i\phi_x^2}\, ,\\
P_x^{23}&=-\cos\theta_x^2\sin\theta_x^3 e^{-i(\phi_x^1+\phi_x^5)} \\
&\hspace{20pt}-\sin\theta_x^1\sin\theta_x^2\cos\theta_x^3 e^{i(\phi_x^2-\phi_x^3+\phi_x^4)}\, ,  \\
P_x^{31}&=-\sin\theta_x^1\cos\theta_x^2\sin\theta_x^3 e^{i(\phi_x^1-\phi_x^3+\phi_x^5)} \\
&\hspace{20pt} -\sin\theta_x^2\cos\theta_x^3 e^{-i(\phi_x^2+\phi_x^4)}\, ,\\
P_x^{32}&=\cos\theta_x^1\sin\theta_x^3 e^{i\phi_x^5}\, ,\\
P_x^{33}&=\cos\theta_x^2\cos\theta_x^3 e^{-i(\phi_x^1+\phi_x^2)} \\
&\hspace{20pt} -\sin\theta_x^1\sin\theta_x^2\sin\theta_x^3 e^{-i(\phi_x^3-\phi_x^4-\phi_x^5)}\, ,
    \end{split}\label{eq:su3-sing-site-param}
\end{equation}
in terms of the three zenith angles $0\leq\theta^1_x,\theta^2_x,\theta^3_x\leq\pi/2$ and the five azimuthal angles $0\leq\phi^1_x,\ldots,\phi^5_x\leq2\pi$ for each plaquette. We collect these angles into a variable $\Omega_x = (\theta^1_x,..,\theta^3_x,\phi^1_x,...,\phi^5_x)$ for ease of notation. The Haar measure is related to the measure of $\Omega$ by~\cite{Bronzan:1988wa}
\begin{equation}
\begin{aligned}
dP_x=\frac1{2\pi^5}&\sin\theta_x^1(\cos\theta_x^1)^3\sin\theta_x^2\cos\theta_x^2\sin\theta_x^3\cos\theta_x^3\\
&\times d\theta_x^1\,d\theta_x^2\,d\theta_x^3\,d\phi_x^1\ldots d\phi_x^5\, .\label{eq:su3-measure}
\end{aligned}
\end{equation}
To compute deformed observables from Monte Carlo samples in the matrix representation, an inverse map of \eqref{eq:su3-sing-site-param} is needed and for example can be specified by
\begin{equation}
\begin{aligned}
    \theta_x^1 & = \text{arcsin}(|P_x^{12}|),\\
    \theta_x^2 & = \text{arccos}\left(|P_x^{11}|/\text{cos}(\theta_x^1)\right),\\
    \theta_x^3 & = \text{arccos}\left(|P_x^{22}|/\text{cos}(\theta_x^1)\right),\\
    \phi_x^1 & =  \arg(P_x^{11}),\\
    \phi_x^2 & =  \arg(P_x^{22}),\\
    \phi_x^3 & =  \arg(P_x^{12}),\\
    \phi_x^4 & =  \arg(P_x^{13}),\\
    \phi_x^5 & =  \arg(P_x^{32}).\\
\end{aligned}
\end{equation}

An $SU(3)$ field configuration in $(1+1)$D with open boundary conditions is defined by a collection of angular variables $\Omega_x$ associated all plaquettes $P_x$ on the lattice.
The $a^{\text{th}}$ component of the deformed angles at site $x$ is denoted by $(\widetilde{\Omega}_x)_a$, and for vertical deformations expanded in Fourier modes is specified by
\begin{equation} \label{eq:su3-deform}
\begin{aligned}
    \tilde{\phi}^a_x &= \phi^a_x + i \kappa_0^{x;\phi^a} + i \sum_{y \leq x} f_{\phi^a}(\Omega_y ; \kappa^{xy}, \lambda^{xy}, \chi^{xy}, \zeta^{xy}), \\
    \tilde{\theta}^a_x &= \theta^a_x + i \sum_{y \leq x} f_{\theta^a}(\Omega_y ; \kappa^{xy},  \lambda^{xy}, \chi^{xy}).
\end{aligned}
\end{equation}
where
\begin{widetext}
\begin{equation} \label{eq:su3-euler-deform}
\begin{aligned}
f_{\theta^a} & = \sum_{m=1}^{\Lambda}{ \kappa_{m}^{xy;a } \text{sin}(2 m \theta^a_y)\bigg\{ 1+\sum_{n=1}^{\Lambda}\big[\sum_{\substack{r\neq a \\ r=1}}^{3}{\lambda_{m n}^{xy;a r}\text{sin}(2 n \theta^r_y)} + \sum_{s=1}^{5}{\eta_{m n}^{xy;a s} \text{sin}(n \phi_y^s+\chi_{m n}^{xy;a s})}\big] \bigg\}}, \\
f_{\phi^a} & = \sum_{m=1}^{\Lambda}{ {{\kappa}}_{m}^{xy;a } \text{sin}(m \phi^a_y + {\zeta}_m^{xy;a})\bigg\{ 1+\sum_{n=1}^{\Lambda}\big[\sum_{r=1}^{3}{{\lambda}_{m n}^{xy;a r}\text{sin}(2  n \theta^r_y)} + \sum_{\substack{s\neq a \\ s=1}}^{5}{{\eta}_{m n}^{xy;a s} \text{sin}(n \phi^s_y+{\chi}_{m n}^{xy;a s})}\big] \bigg\}}.
\end{aligned}
\end{equation}
\end{widetext}
Deformed observables analogous to Eq.~\eqref{eq:su2-deformed-obs} can be constructed for the $SU(3)$ case using this parameterization and ratios of the deformed and undeformed Haar measure factors obtained from Eq.~\eqref{eq:su3-measure}.

\subsection{Results}

\begin{figure*}[t]
    \centering
    \includegraphics{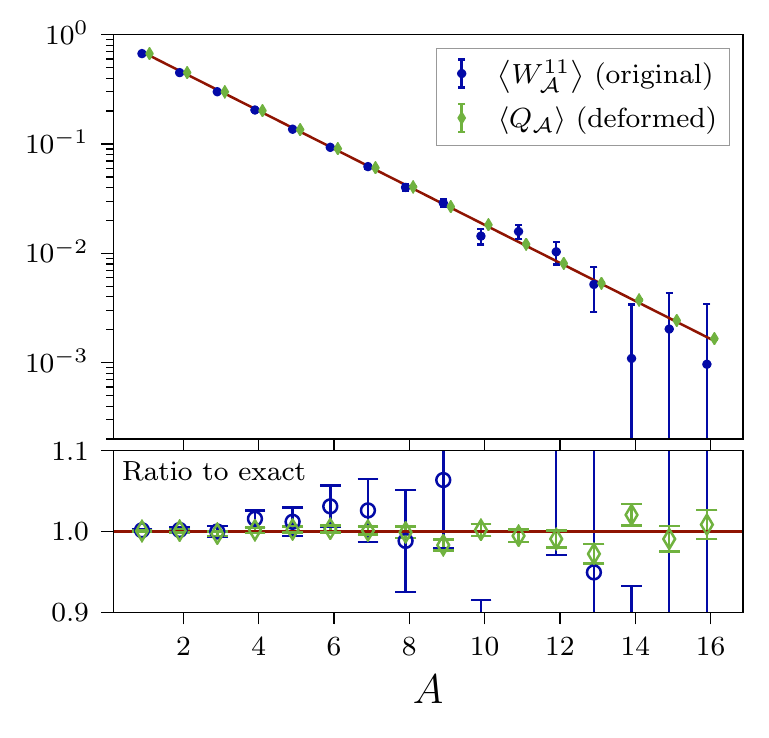}
    \includegraphics{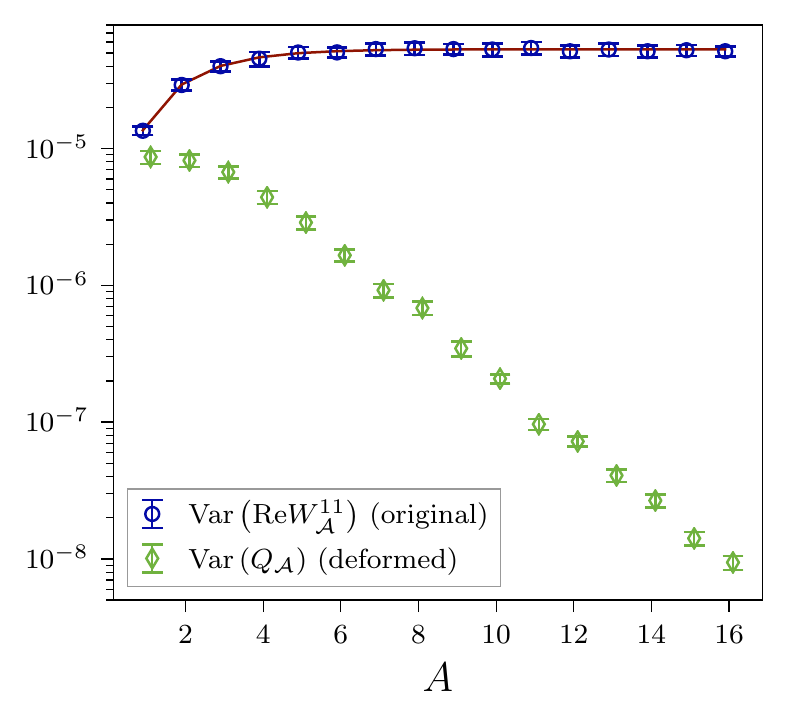}
    \includegraphics{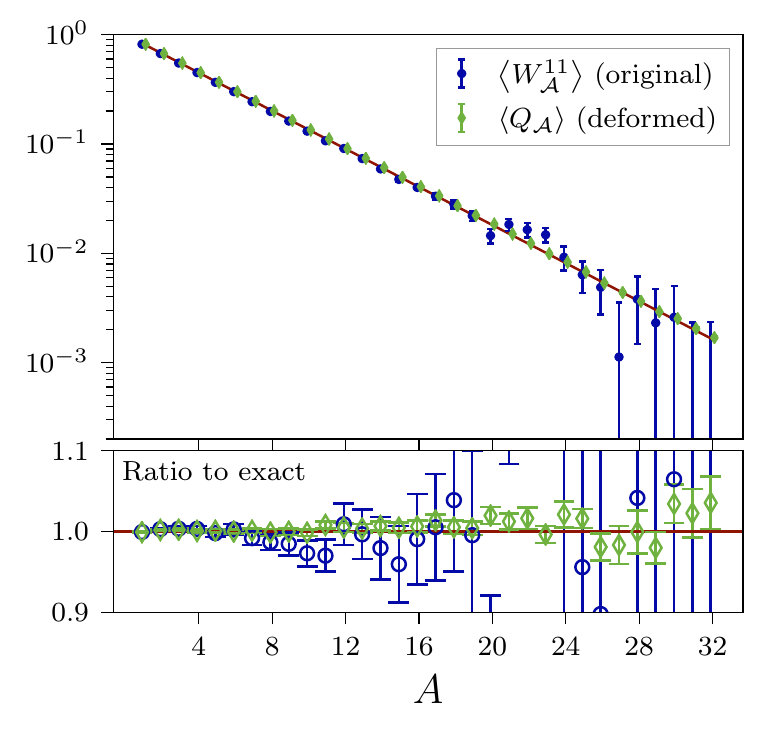}
    \includegraphics{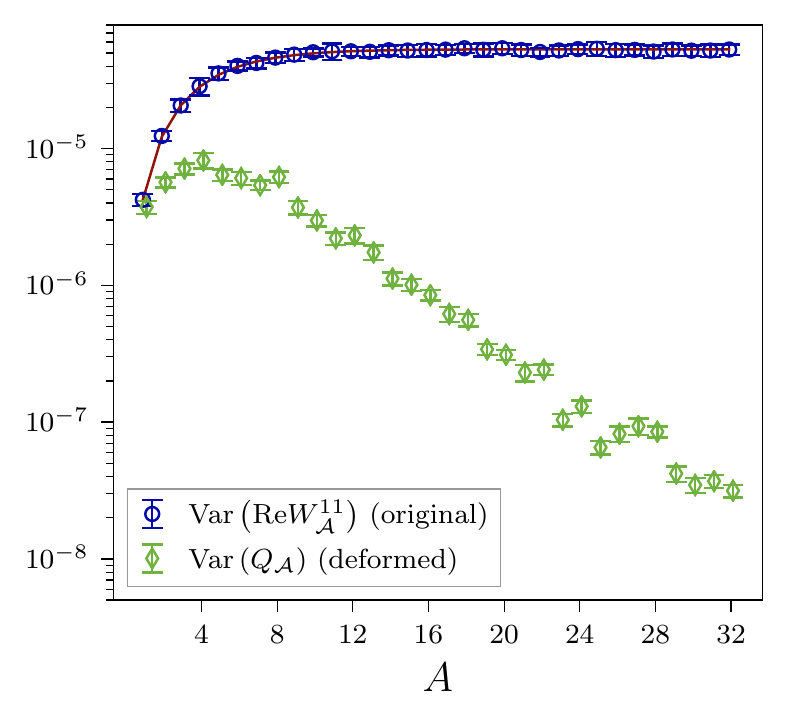}
    \includegraphics{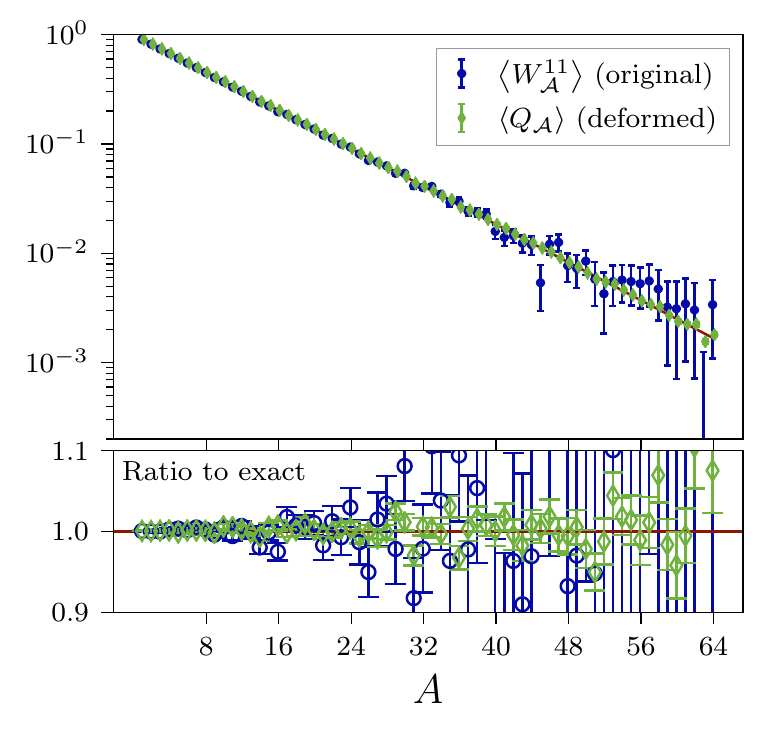}
    \includegraphics{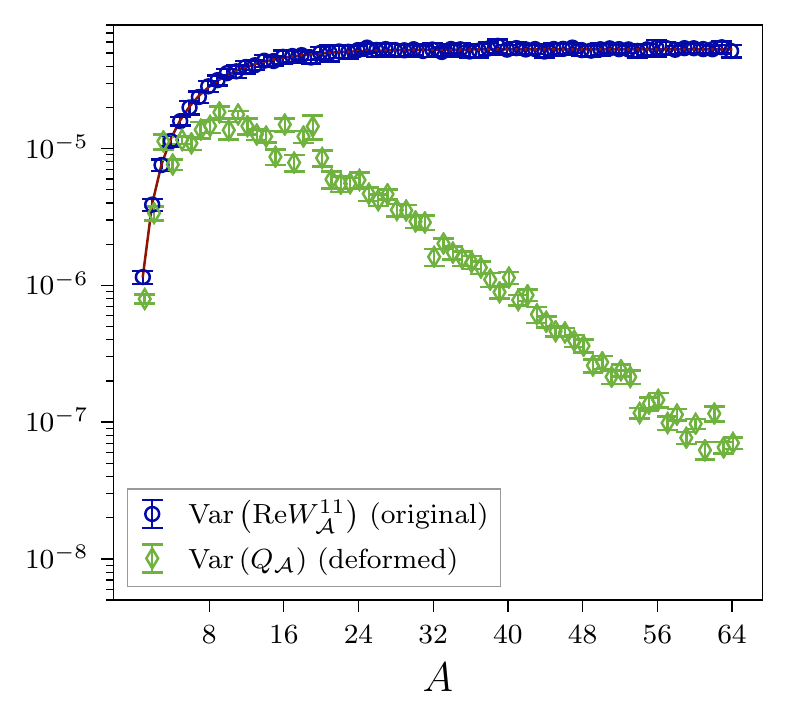}
    \caption{$SU(3)$ Wilson loop expectation values and variances for ensembles with three different values of the gauge coupling that from top to bottom correspond to $g=0.72,\ 0.53,\ 0.38$. Red lines correspond to analytical results for $\left<W_{\mathcal{A}}^{11}\right> = \left< \tr(W_{\mathcal{A}})/3 \right>$ (left plots) and $\Var(\Re W_{\mathcal{A}}^{11})$ (right plots).}
    \label{fig:su3_Euler_observable}
\end{figure*}

Practical performance of these deformations was investigated by optimizing Wilson loop variance using the three sets of $SU(3)$ parameters detailed in Table~\ref{tab:couplings} as in the $SU(2)$ case. The couplings were tuned to match the string tensions used for $SU(2)$ gauge theory and correspond to lattice spacings varying by a factor of 4. For each choice of parameters, an ensemble of $n = 32000$ configurations was generated using the factorized HMC method discussed in Sec.~\ref{sec:su2-mc}. Fig.~\ref{fig:su3_Euler_observable} shows variance reduction for Wilson loops of all possible sizes for the three lattice spacings studied. At the largest loop areas, we found variance reduction of greater than three orders of magnitude. Across all three ensembles, analytical results for the Wilson loop expectation values and variances were reproduced by the undeformed Monte Carlo estimates. The expectation value of the deformed observable is consistent with the analytical and original Monte Carlo results, while the variance of the deformed observable exponentially decreases with increasing $\sigma A$.

\begin{figure}[t]
    \centering
    \includegraphics{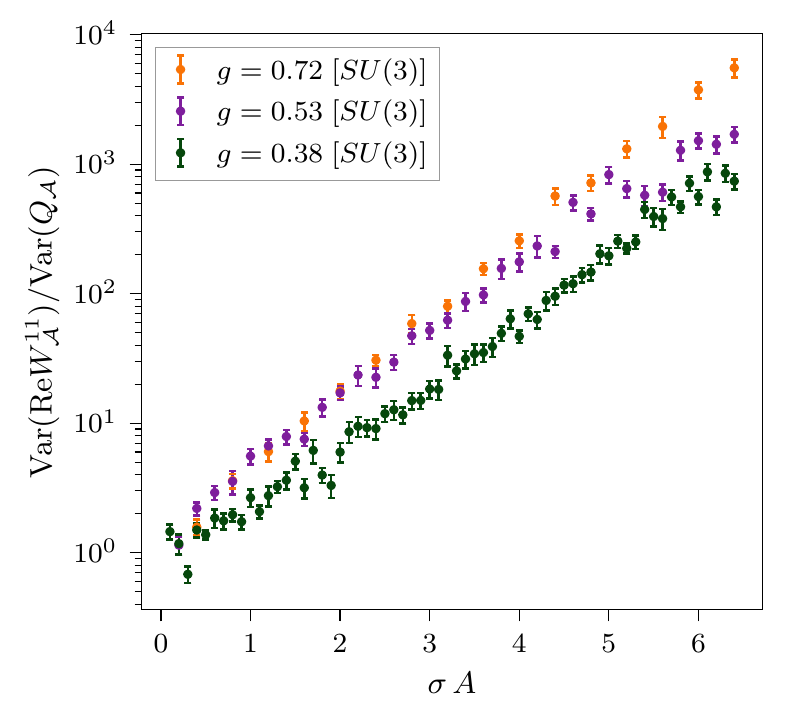}
    \caption{$SU(3)$ Wilson loop variance ratios of standard observables to deformed observables for ensembles with three different values of the gauge coupling that correspond to $g=0.72,\ 0.53,\ 0.38$ (top to bottom).}
    \label{fig:su3_Euler_lattice}
\end{figure}

Fig.~\ref{fig:su3_Euler_lattice} compares the variance reduction achieved at all three lattice spacings versus physical loop area $\sigma A$. We found approximately equivalent improvement in the variance at the two coarser lattice spacings ($g=0.72$ and $g=0.53$), and a small, yet significant, decrease in the variance improvement achieved at the finest lattice spacing ($g=0.38$). Despite this, the variance was reduced by three orders of magnitude at the largest area on the finest lattice by using an optimized deformed observable, and at all three couplings variance improvements are consistent with exponential in the physical loop area.
The number of parameters to be optimized grows with the volume in lattice units, and the analogous results observed for $SU(2)$ gauge theory suggest that the results at finer lattice spacings could be partially explained by increased difficulty in training the larger number of parameters.

\begin{figure}[t]
    \centering
    \includegraphics{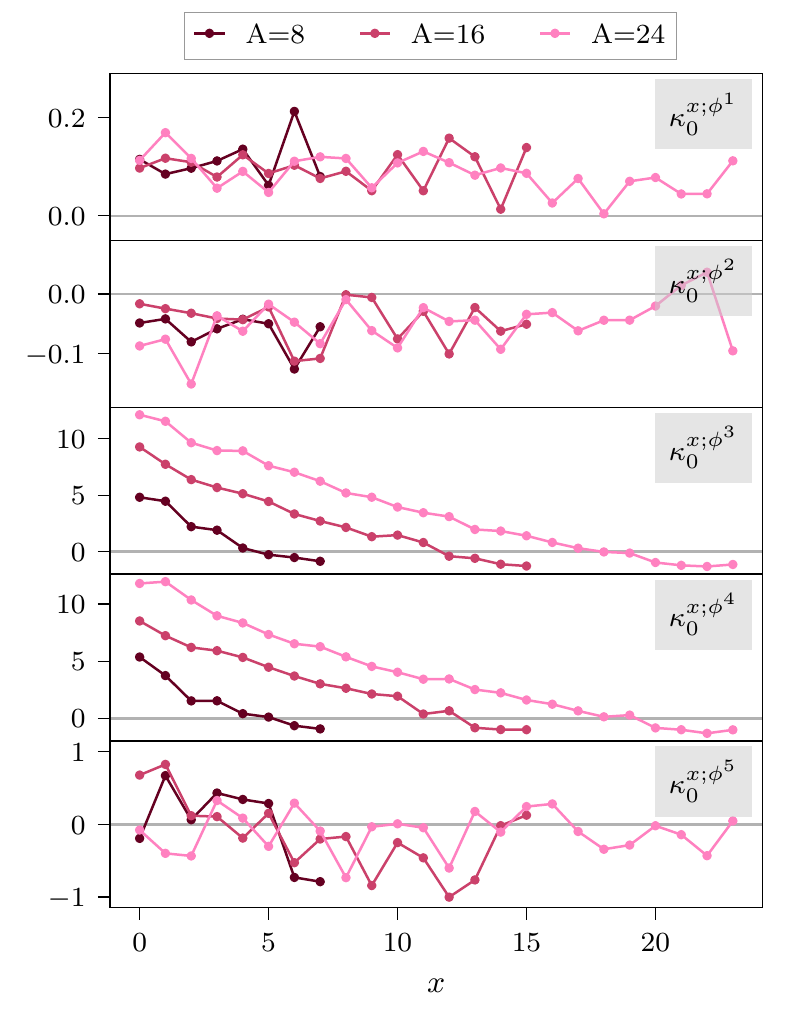}
    \caption{The parameters of the optimal manifold for Wilson loop $Q_{\mathcal{A}}$ at three different areas $A$, as determined on the ensemble with total volume $V=32$ and $g=0.53$. Optimization for each $Q_{\mathcal{A}}$ was initialized using the parameters for optimal $Q_{\mathcal{A}'}$ with region $\mathcal{A}'$ of area $A-1$, as detailed in the main text.}
    \label{fig:su3_Euler_manifold}
\end{figure}

The $\Lambda = 0$ manifold parameterization involves few enough parameters that it is possible to investigate the structure of the optimal parameters similarly to the case of $SU(2)$ gauge theory. As shown in Fig.~\ref{fig:su3_Euler_manifold}, we found that the optimized values of $\kappa_0^{x;3}$ and $\kappa_0^{x;4}$ decrease with $x$, while $\kappa_0^{x;1}$ and $\kappa_0^{x;2}$ appear to converge towards approximately constant positive and negative values, respectively. The final parameter $\kappa_0^{x;5}$ fluctuates in both the positive and negative directions. These results can be qualitatively explained by expanding the $(1,1)$ component of the Wilson loop for small area. For $A=1$ the Wilson loop is equivalent to $P_x$, for which the $(1,1)$ component is
\begin{equation}
P_x^{11} = \cos\theta^1_x \cos\theta^2_x e^{i \phi^1_x}.
\end{equation}
The magnitude of this quantity can be reduced by shifting $\phi^1 \rightarrow \widetilde{\phi}_x^1 = \phi_x^1 + i \lambda$, which is consistent with the positive $\kappa_0^{x;1}$ values obtained after optimization shown in Fig.~\ref{fig:su3_Euler_manifold}.
Extending the analysis to the $A=2$ Wilson loop, the $(1,1)$ component is given by
\begin{equation}
\begin{aligned}
& (P_x P_{x'})^{11} = \\
&\quad e^{i \phi^1_x + i \phi^{1}_{x'}} \cos\theta^1_x \cos\theta^{1}_{x'} \cos\theta^2_x \cos\theta^{2}_{x'} \\
&\quad - e^{i \phi^{1}_{x'} + i \phi^{2}_{x'} + i(\phi^3_x - \phi^{3}_{x'})} \cos\theta^{2}_{x'} \cos\theta^{3}_{x'} \sin\theta^1_x \sin\theta^{1}_{x'} \\
&\quad - e^{-i \phi^{2}_{x'} + i (\phi^4_x - \phi^{4}_{x'})} \cos\theta^1_x \cos\theta^{3}_{x'}\sin\theta^{2}_x \sin\theta^{2}_{x'}\\
&\quad - e^{i \phi^4_x + i \phi^{1}_{x'} - i\phi^{3}_{x'} + i \phi^{5}_{x'}} \cos\theta^1_x \cos\theta^{2}_{x'} \sin\theta^{1}_{x'} \sin\theta^2_x \sin\theta^{3}_{x'} \\
&\quad + e^{i \phi^3_x - i \phi^{4}_{x'} - i \phi^{5}_{x'}} \sin\theta^1_x \sin\theta^{2}_{x'} \sin\theta^{3}_{x'} \, .
\end{aligned}
\end{equation}
The magnitude of the first, second, and fourth terms are reduced by shifting $\phi^1_x \rightarrow \phi^1_x + i \lambda$ and $\phi^{1}_{x'} \rightarrow \phi^1_{x'} + i \lambda$ with $\lambda > 0$. The magnitude of the second and third terms can be reduced by shifting $\phi^3_x - \phi^{3}_{x'} \rightarrow \phi^3_x - \phi^{3}_{x'} + i \delta$ and $\phi^4_x - \phi^{4}_{x'} \rightarrow \phi^4_x - \phi^{4}_{x'} + i \delta$ with $\delta > 0$; this is also consistent with a positive imaginary shift of $i \delta$ in $\phi^3_x - \phi^{4}_{x'}$ and $\phi^4_x - \phi^{3}_{x'}$, reducing the magnitude of the fourth and fifth terms. These deformations result in reduced magnitude and correspondingly lower variance. Deformations with these qualitative features are reproduced in the optimized manifolds found for $\Lambda = 0$. Finally, we note that $\phi^{2}_{x'}$ appears in the exponent with opposite signs in the second and third terms, and similarly for $\phi^{5}_{x'}$ in the fourth and fifth terms, so there is no constant vertical deformation of these terms that will reduce the overall magnitude.

\begin{figure}[t]
    \centering
    \includegraphics{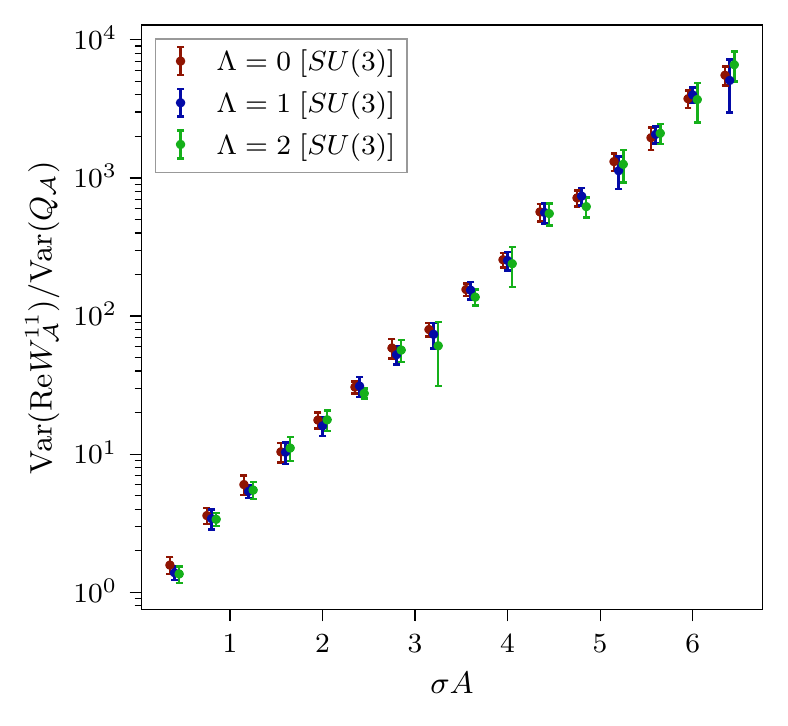}
    \caption{$SU(3)$ Wilson loop variance ratios of standard observables to deformed observables for ensembles with $g=0.53$ and three different values of the manifold parameterization cutoff.}
    \label{fig:su3_Euler_cutoff}
\end{figure}

Analogously to the case of $SU(2)$ gauge theory, no significant StN improvements were obtained by including higher Fourier modes. Fig.~\ref{fig:su3_Euler_cutoff} directly compares optimized manifolds for $\Lambda \leq 2$ at the coarsest lattice spacing, showing that the variance improvement was unchanged by including higher Fourier modes.
Constant vertical deformations therefore appear to saturate the StN benefits achieved by general Fourier series parameterizations for deformed Wilson loops in $(1+1)$D.
More complicated parameterizations of contour deformations may prove more useful in $(3+1)$D; however, numerical studies of deformed observables in $(3+1)$D lattice gauge theory are left to future work.

\section{Conclusions}\label{sec:cons}

In this work, we have defined a family of complex manifolds for path integral deformations in $SU(N)$ lattice field theories. The manifolds introduced here are described in terms of an angular parameterization of each $SU(N)$ variable, with dependence between variables at differing spacetime sites restricted to enforce a triangular Jacobian. 
We find that choosing a parameterization in which the observable of interest can be written as $\obs = e^{i \theta} X$ is a useful practical choice, allowing constant shift deformations in the parameter $\theta$ to make substantial progress in reducing noise. Choosing a spacetime dependence of the deformation that gives rise to a triangular Jacobian is key to ensuring the deformed integral can be computed efficiently, as the Jacobian determinant can then be evaluated with cost linear in the number of spacetime lattice sites.

This manifold parameterization can be combined with the method of deformed observables introduced in Ref.~\cite{Detmold:2020ncp} to reduce noise in observables. This approach is applicable when the action is real and the Boltzmann weight $e^{-S}$ can be treated as a probability measure. We stress that this method of deformed observables does not require changing the Monte Carlo sampling, despite being based on an analysis of contour deformation of the entire path integral, and can be thought of as an approach to analytically relate observables with identical expectation values and different variance. Keeping the Monte Carlo weights unchanged allows manifold optimization using estimates of the deformed observable variance computed with respect to a fixed Monte Carlo ensemble. There is a tradeoff between the cost of optimizing manifold parameters and the statistical precision gained. In practice, we find that initializing manifold parameters from optimal parameters for similar observables significantly reduces the associated cost.

This method was shown in Sec.~\ref{sec:su2} and \ref{sec:su3} to improve the variance of Wilson loop observables in $(1+1)$D $SU(2)$ and $SU(3)$ lattice gauge theory by orders of magnitude. 
For the original Wilson loop observables, the signal-to-noise ratio decreases exponentially with area. The deformed observables mitigate this StN problem, and in particular we find that the improvements are consistent with an exponential in the physical Wilson loop area, with the most significant reduction in noise for Wilson loops of the largest area. The improvement in variance was empirically found to be similar at three different lattice spacings, though less of an improvement was seen at the finest lattice spacing for $SU(3)$; the achieved improvements in the continuum limit and for other theories is thus an interesting subject of future investigation. However, we stress that making any gains at finite lattice spacing is still a significant step forward due to the convenience of the method: optimizing a deformation on a fixed ensemble quickly gives new observables that encode the same physical content while having significantly reduced noise.

In demonstrating the method, we focused here on a particular deformation of the angular parameters based on a Fourier series expansion and shift in the imaginary direction only.
Writing the observable phase fluctuations in terms of these periodic angular parameters that can be shifted by a constant in the imaginary direction led to deformed observables with significant StN improvement.
The surprising result that zero-mode terms alone significantly reduce noise, with neither dependence between plaquettes at different spacetime sites nor dependence on the values of the angular parameters themselves, suggests that the majority of the StN problem in these $(1+1)$D theories arises from independent local fluctuations of $SU(N)$ angular parameters.

Complications are expected in higher dimensions, as Gauss' Law implies that plaquettes at differing spacetime locations and orientations must satisfy many independent constraints. Deformations thus cannot independently address fluctuations in each plaquette included in a Wilson loop, or more generally in each fundamental degree of freedom included in an observable. It is therefore an interesting line of future work to determine how best to incorporate this spacetime dependence in higher dimensional applications of path integral contour deformations. The approach employed here is one of many possible approaches to creating expressive transformations with efficiently computable Jacobians. This issue has been explored in some depth in normalizing flows for sampling in many contexts, including image generation and ensemble generation for lattice field theory; see Refs.~\cite{papamakarios2019normalizing,Kobyzev_2020} for recent reviews. Similar techniques may prove more fruitful in future applications of path integral contour deformations to observables of phenomenological interest in $(3+1)$D lattice gauge theories.

\begin{acknowledgments}
W.D.\ and G.K.\ are supported in part by the U.S.\ DOE grant No.\ DE-SC0011090.
W.D.\ is also supported by the SciDAC4 award DE-SC0018121. H.L.\ is supported by a Department of Energy QuantiSED grant. N.C.W.\ is supported by the U.S.\ DOE under Grant No.\ DE-FG02-00ER41132.
This manuscript has been authored by Fermi Research Alliance, LLC under Contract No.\ DE-AC02-07CH11359 with the U.S. Department of Energy, Office of Science, Office of High Energy Physics.
\end{acknowledgments}

\appendix

\section{Single-variable \texorpdfstring{$SU(N)$}{SU(N)} integrals} \label{app:single-site-integrals}

The calculation of $z$ for $SU(N)$ can be performed analogously to the $U(N)$ case considered in Refs.~\cite{Gross:1980he,Wadia:2012fr} with further details for the $SU(N)$ case given in Ref.~\cite{Drouffe:1983fv}.
In the eigenbasis of $P$, the Haar measure is given by
\begin{equation}
   \begin{split}
      dP &= \frac{1}{N!} \prod_{I=1}^N \left[ \frac{d\theta_I}{2\pi} \prod_{J < I} \left| e^{i\theta_I} - e^{i\theta_J} \right|^2 \right. \\
      &\hspace{20pt} \left. \times \sum_{n=-\infty}^\infty 2\pi \delta\left( \sum_{K=1}^N \theta_K + 2\pi n \right) \right],
   \end{split}\label{eq:eigenHaar}
\end{equation}
where the $\delta$-function enforces the unit determinant condition of $SU(N)$.
Using the Fourier series representation of the $\delta$-function for the compact variable $\sum_{K=1}^N \theta_K$, this can be expressed as
\begin{equation}
   \begin{split}
      dP = \frac{1}{N!} \prod_{I=1}^N \left[ \frac{d\theta_I}{2\pi} \prod_{J < I} \left| e^{i\theta_I} - e^{i\theta_J} \right|^2 \sum_{q=-\infty}^\infty e^{i q \theta_I} \right].
   \end{split}\label{eq:eigenHaar2}
\end{equation}
The product of $e^{i\theta_I} - e^{i\theta_J}$ factors can be expressed in terms of the determinant of a Vandermonde matrix~\cite{Gross:1980he,Wadia:2012fr}. 
The $SU(N)$ Haar measure is given by a sum of similar determinants, and $z$ can be expressed as~\cite{Drouffe:1983fv}
\begin{equation}
   \begin{split}
      z = \sum_{q=-\infty}^\infty \det(\mathcal{Z}^q),
   \end{split}\label{eq:eigenHaar3}
\end{equation}
where the entries of the matrix $\mathcal{Z}^q$ are given by
\begin{equation}
   \begin{split}
      \mathcal{Z}^q_{IJ} &\equiv \int \frac{d\theta}{2\pi} e^{i [q + I - J]\theta} e^{\frac{2}{g^2}\cos(\theta)} \\
      &= I_{q+I-J}\left(\frac{2}{g^2}\right),
   \end{split}\label{eq:Zmat}
\end{equation}
where $I_n(x)$ is a modified Bessel function.
For example, in the $SU(2)$ case $z$ is given explicitly by
\begin{equation}
   \begin{split}
      z^{SU(2)} &= \sum_{q=-\infty}^{\infty} \left[I_q\left(\frac{2}{g^2}\right)\right]^2 - I_{q+1}\left(\frac{2}{g^2}\right)I_{q-1}\left(\frac{2}{g^2}\right).
   \end{split}\label{eq:z2}
\end{equation}
A simpler but equivalent form $z^{SU(2)} = \frac{g^2}{2} I_1(4/g^2)$ can also be derived using the parameterization introduced in Section~\ref{sec:su2}.  From this, we can derive an expression for $\avg{\chi_1}$ from taking derivatives of $z$
\begin{equation}
\label{eq:avgchi}
    \frac{\partial}{\partial(2N/g^2)}\text{log} \, z = -\frac{g^3}{4 N}\frac{\partial}{\partial g}\text{log} \, z = \avg{\chi_1} = \text{e}^{-\sigma}.
\end{equation}

\begin{table}
\centering
\begin{tabular}
{>{\centering}m{1cm} @{\hspace{0.75cm}} >{\centering}m{1.5cm} >{\centering\arraybackslash}m{2.5cm}}
\toprule
$r$& $d_r$ & $\chi_r$\\
\midrule
$\{1\}$ & $N$ & $\tr U $  \\[0.2em]
$\{2\}$ & $\frac{N(N+1)}{2}$ &$\frac{1}{2}\left(\tr^2U+\tr U^2\right) $  \\[0.2em]
$\{1,1\}$ & $\frac{N(N-1)}{2}$ &$\frac{1}{2}\left(\tr^2 U-\tr U^2\right) $\\[0.2em]
$\{1,-1\}$ & $N^2-1 $  &$\left|\tr U\right|^2-1 $\\
\bottomrule
\end{tabular}
\caption{\label{tab:vrnum} Properties of group representations: the dimension $d_r$ and the character $\chi_r$. We have followed the normalizations in Table 14 of Ref.~\cite{Drouffe:1983fv}.}
\end{table}

Similar to $\left< \tr(W_{\mathcal{A}})/N \right>$, we can factorize $\left< \left| \tr(W_{\mathcal{A}})^2 \right| / N^2 \right>$ and $\left< \tr(W_{\mathcal{A}})^2 / N^2 \right>$ into integrals of character functions over single-elements of $SU(N)$ using an identity~\cite{Creutz:1978ub} 
\begin{equation}
   \begin{split}
      &\int d\Omega\  \Omega_{i_1j_1} \Omega^\dagger_{k_1l_1} \Omega_{i_2j_2} \Omega^\dagger_{k_2l_2}  \\
      &= \frac{1}{N^2-1} \left( \delta_{i_1l_1} \delta_{j_1k_1} \delta_{i_2l_2} \delta_{j_2k_2} + \delta_{i_1l_2} \delta_{j_1k_2} \delta_{i_2l_1} \delta_{j_2k_1} \right)  \\
      & \hspace{10pt} - \frac{1}{N(N^2-1)} \left( \delta_{i_1l_2} \delta_{j_1k_1} \delta_{i_2l_1} \delta_{j_2k_2} + \delta_{i_1l_1} \delta_{j_1k_2} \delta_{i_2l_2} \delta_{j_2k_1} \right).
   \end{split}\label{eq:SUNidvar1}
\end{equation}
From Table~\ref{tab:vrnum}, we recognize that $\left< \left| \tr(W_{\mathcal{A}})^2 / N^2 \right| \right>$ is related to $\chi_{1,-1}(P_x)$. Thus, applying Eq.~\eqref{eq:SUNidvar1} we can derive that
\begin{equation}
   \begin{split}
      &\int d\Omega \ \chi_{1,-1}(A\Omega P \Omega^\dagger B)\\
      &= \tr(AB)\tr(B^\dagger A^\dagger)\langle\chi_{1,-1}\rangle\\
      &\qquad+ \tr(A A^\dagger B^\dagger B)\frac{1}{N}\left(1 - \langle\chi_{1,-1}\rangle\right)-1 \\
      &= \left[\tr(AB)\tr(B^\dagger A^\dagger)-1\right]\langle\chi_{1,-1}\rangle\\
      &= \chi_{1,-1}(AB)\langle\chi_{1,-1}\rangle
   \end{split}\label{eq:varfac1}
\end{equation}
where
\begin{equation}
   \begin{split}
      \langle\chi_{1,-1}\rangle &\equiv \frac{1}{z} \int dP \ \frac{1}{N^2-1}\  \chi_{1,-1}(P) \ e^{\frac{1}{g^2} \tr \left( P + P^\dagger \right)}.
   \end{split}\label{eq:w2def}
\end{equation}
Iterating this identity within $\left< \left| \tr(W_{\mathcal{A}})^2 \right| \right>$ gives
\begin{equation}
   \begin{split}
      &\left< \left| \tr(W_{\mathcal{A}})^2 \right| \right> \\
      &\quad = N^2 \langle\chi_{1,-1}\rangle^A +  (1-\langle\chi_{1,-1}\rangle)\sum_{k=0}^{A-1} \left( \frac{\langle\chi_{1,-1}\rangle - 1}{N^2} \right)^k  \\
      &\quad = 1+(N^2 - 1)\langle\chi_{1,-1}\rangle^A.
   \end{split}\label{eq:var1}
\end{equation}
Since $\tr(W_{\mathcal{A}})^2$ is not a character $\chi_r$, attempting to factorize it generates new terms.  Specifically, in addition to $\tr(W_{\mathcal{A}^\prime})^2$ one finds $\tr(W_{\mathcal{A}^\prime}^2)$ for $\mathcal{A}^\prime \subset \mathcal{A}$. Instead, a basis involving $\chi_2(P)$ and $\chi_{1,-1}(P)$ can be constructed from linear combinations of these traces and satisfies 
\begin{equation}
\int d\Omega\ \chi_2(A \Omega P \Omega^\dagger B)=\frac{2}{N(N+1)}\chi_2(AB)\avg{\chi_2},\\
\label{eq:varfac2}
\end{equation}
and
\begin{equation}
\int d\Omega\ \chi_{1,1}(A \Omega P \Omega^\dagger B)=\frac{2}{N(N-1)}\chi_{1,1}(AB)\avg{\chi_{1,1}},\\
\label{eq:varfac3}
\end{equation}
where as in Eq.~\eqref{eq:w2def} we have factorized using the expectation values of the characters
\begin{equation}
   \begin{split}
      \avg{\chi_2} &\equiv \frac{1}{z} \int dP \ \frac{2}{N(N+1)}\  \chi_{2}(P) \ e^{\frac{1}{g^2} \tr \left( P + P^\dagger \right)}, \\
      \avg{\chi_{1,1}} &\equiv \frac{1}{z} \int dP \ \frac{2}{N(N-1)}\  \chi_{1,1}(P)\ e^{\frac{1}{g^2} \tr \left( P + P^\dagger \right)}.
   \end{split}\label{eq:w2pdef}
\end{equation}
Putting these together and iterating gives
   \begin{equation}
      \begin{split}
         \avg{ \tr(W_{\mathcal{A}})^2 } &= \frac{N(N+1)}{2}\avg{\chi_2}^A+\frac{N(N-1)}{2}\avg{\chi_{1,1}}^A.
      \end{split}\label{eq:var2}
   \end{equation}
Combining Eqs.~\eqref{eq:var1} and~\eqref{eq:var2} gives the general expression for variance of the $SU(N)$ Wilson loop
\begin{equation}
\begin{split}
    & \Var[\Re \tr(W_{\mathcal{A}}^{SU(N)})/N] \\
    &= \frac{1}{2N^2} \left< \left| \tr(W_{\mathcal{A}})^2 \right| \right> + \frac{1}{2N^2} \left< \tr(W_{\mathcal{A}})^2 \right> \\
    &\qquad\qquad - \frac{1}{N^2} \left< \tr(W_{\mathcal{A}}) \right>^2 \\
    &=\frac{1}{2N^2}\bigg[1+(N^2 - 1)\langle\chi_{1,-1}\rangle^A +\frac{N(N+1)}{2}\avg{\chi_2}^A
    \\&\qquad\qquad+\frac{N(N-1)}{2}\avg{\chi_{1,1}}^A\bigg]- e^{-2 \sigma^{SU(N)}A}.
\end{split}
\end{equation}
The character expectation values can be obtained for general $SU(N)$ gauge groups by numerically evaluating the integrals in Eq.~\eqref{eq:w2pdef}.

For $SU(2)$, it is straightforward to analytically evaluate the character expection values appearing in Eq.~\eqref{eq:w2pdef}. Not all representations of $SU(2)$ are independent using the enumeration in Table~\ref{tab:vrnum}, and in particular $\chi_{2}(P)=\chi_{1,-1}(P)$ and $\chi_{1,1}(P)=\chi_0(P)=1$. After removing the redundant characters, the remain integrals $\chi_{j}(P)$ can be solved using expression of the characters in terms of angles~\cite{Drouffe:1983fv} 
\begin{equation}
 \chi_j=\frac{\sin([j+1/2]\alpha)}{\sin(\alpha/2)}
\end{equation}
where $j=0,\frac{1}{2},1\cdots$ index the unique characters of $SU(2)$ and $r=\{1,-1\}$ equals $j=1$.  With these, one has 
\begin{equation}
 \frac{1}{z}\int DP \frac{1}{d_j}\chi_j(P)\, e^{\frac{1}{g^2}\Tr(P+P^\dag)}=\frac{I_{2j+1}(4/g^2)}{I_1(4/g^2)}.
\end{equation}
From which we derive
\begin{equation}
   \begin{split}
      \text{Var} & [\Re \tr( W_{\mathcal{A}}^{SU(2)} ) / N] \\
      &= \frac{1}{4} + \frac{3}{4}\left( \frac{I_3(4/g^2)}{I_1(4/g^2)} \right)^A - e^{-2\sigma^{SU(2)} A},
   \end{split}
\end{equation}

\section{Alternative \texorpdfstring{$SU(2)$}{SU(2)} coordinates}\label{app:su2-euler}

\begin{figure}
    \centering
    \includegraphics{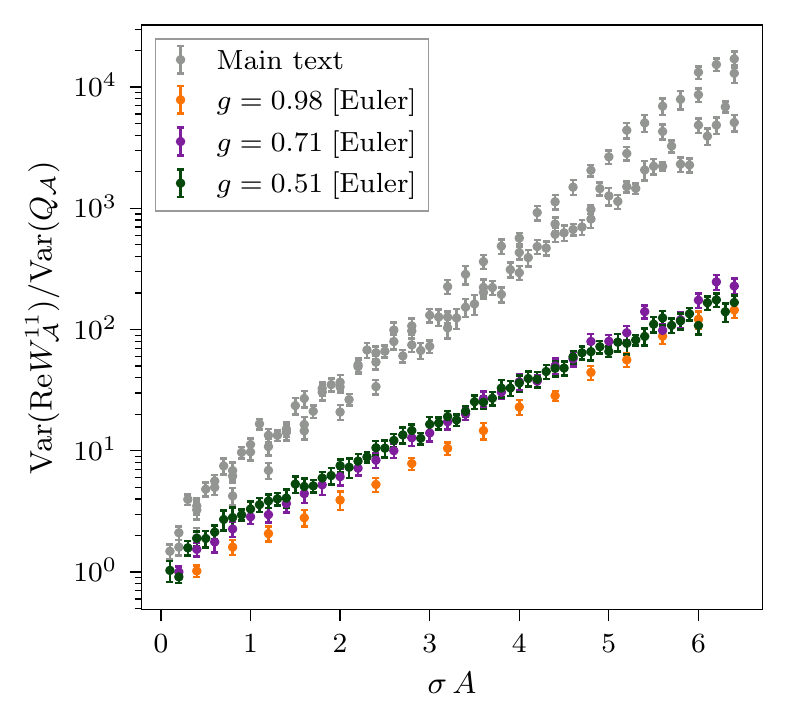}
    \caption{Colored points $SU(2)$ Wilson loop variance ratios using the alternative gauge field parameterization defined in Eq.~\eqref{eq:euler-link-param}. Gray points show analogous variance ratios using the parameterization defined in Sec.~\ref{sec:su2-param} for comparison and are identical to the results in Fig.~\ref{fig:su2_Euler_lattice}. 
    \label{fig:su2_Euler_vs_oldEuler_lattice} }
\end{figure}

As an alternative to the parameterization presented in Sec.~\ref{sec:su2-param}, plaquettes $P_x \in SU(2)$ can be represented as
\begin{equation}
\begin{split}
   P_x &= \exp \left( \frac{i \alpha_x}{2} \hat{n}_x\cdot \vec{\sigma} \right) = \cos\left( \frac{\alpha_x}{2} \right) + i \sin\left(\frac{\alpha_x}{2} \right) \hat{n}_x \cdot \vec{\sigma}.
    \end{split}
\label{eq:euler-link-param}
\end{equation}
The $\mathfrak{su}(2)$ unit vector $\hat{n}_{x}$ can be further parameterized as
\begin{equation} \label{eq:nhat-param}
    \hat{n}_x = (\cos\phi_x \sin\theta_x,\ \sin\phi_x \sin\theta_x,\ \cos\theta_x),
\end{equation}
and a general $SU(2)$ group element $P_{x}$ can be parameterized in terms of the three angles
\begin{equation}
\begin{split}
    0 \leq \alpha_x < 2\pi, \hspace{20pt}
    0 \leq \phi_x < 2\pi,\hspace{20pt} 0 \leq \theta_x < \pi.
    \end{split}
\end{equation}
The Haar measure is given in these coordinates as
\begin{equation}
\begin{split}
dP_x &= \frac{1}{4\pi^2} \sin^2\left(\frac{\alpha_x}{2}\right) d\alpha_x \sin\theta_x d\theta_x d\phi_x,
\end{split}
\end{equation}
The inverse map needed to obtain these angular parameters for an $SU(2)$ matrix $P_x$ is given by
\begin{equation}
   \begin{split}
      \alpha_x &= 2\ \text{arccos}\left[ \frac{1}{2} \left( P_x^{11} +  P_x^{22} \right) \right] \\
      \theta_x &= \text{arccos}\left[ \frac{P_x^{11} -  P_x^{22}}{2i\sin(\alpha_x/2)} \right] \\
      \phi_x &= \frac{1}{2}\text{arg}\left[ \frac{P_x^{21}}{P_x^{12}} \right].
   \end{split}\label{eq:su2angledef}
\end{equation}
As with Eq.~\eqref{eq:su2-param-bwd}, these are not entire functions of $P_x$ but this is not an obstacle for contour deformation because it is only the parameterization given by Eqs.~\eqref{eq:euler-link-param} and \eqref{eq:nhat-param} that determines whether path integrands can be interpreted as holomorphic functions of the angles $\{\alpha_{x},\ \theta_{x},\ \phi_{x}\}$ associated with $P_x$.

Deformed observables starting with the $(1,1)$ component of $SU(2)$ Wilson loops can be defined using this parameterization.
A family of vertical deformations for $\{\alpha_{x},\ \theta_{x},\ \phi_{x}\}$ can be defined analogously to the deformation described in Sec.~\ref{sec:su2-fourier}.
Since $\alpha_{x}$ and $\theta_{x}$ have fixed (non-identified) integration contour endpoints, a constant vertical deformation can only be applied to $\phi_{x}$.
For $A=1$ in particular, $\Tr(P_x) = \cos(\alpha_x/2)$, and the trace is independent of the only constant vertical deformation that can be applied.
Neither this constant vertical deformation nor non-constant vertical deformations corresponding to Fourier basis cutoffs $\Lambda = 1,2$ lead to statistically significant variance reduction with $A=1$.
As shown in Fig.~\ref{fig:su2_Euler_vs_oldEuler_lattice}, for $A>1$ deformed observable results using this parameterization with $\Lambda = 0$ do lead to significant variance reduction when compared to undeformed contour results.
However, orders of magnitude less variance reduction is obtained for large area Wilson loops using optimized deformed observables with this parameterization when compared to results using the parameterization explored in Sec.~\ref{sec:su2-param}.
The fact that  the parameterization in Eq.~\eqref{eq:euler-link-param} leads to less variance reduction than the parameterization in Sec.~\ref{sec:su2-param} can be intuitively explained by the inability of constant vertical deformations to decrease the magnitudes of the $(1,1)$ components of (products of) $SU(2)$ matrices using the parameterization Eq.~\eqref{eq:euler-link-param}.
The significance of the difference between the results demonstrates the utility of rewriting observables before deformation in achieving practical StN improvements, as discussed in Sec.~\ref{sec:rewriting}.

\section{Regularization terms to avoid overtraining and overlap problems} \label{sec:loss-reg}

When $\Re{\tilde{S}}$ is significantly different from $S$, we can encounter an overlap problem for training and evaluation; both processes involve factors of $e^{-\Re{\tilde{S}} + S}$ that can have very large magnitude fluctuations in this situation. To mitigate this problem, it is helpful to include regularization terms in the loss function $\mathcal{L}$. These terms may bias the exact loss minimum away from the optimal, but allow closer convergence to that optimal solution given finite statistics estimates of $\mathcal{L}$. The strength of these terms is controlled by a small parameter $\epsilon$.

We discuss two possible terms here. First, an L2 regularizer~\cite{AndersWeightDecay} may be used, which simply ensures the parameters controlling the deformation all remain close to zero. Generically labeling those parameters as $\lambda_i$, this loss term can be written
\begin{equation}
    \mathcal{L}_{\text{L2}} \equiv \epsilon \sum_i \left| \lambda_i \right|^2.
\end{equation}
In the limit of $\epsilon \rightarrow \infty$, the parameters $\lambda_i$ are forced to zero and the optimization procedure must remain at the original manifold. A smaller choice of $\epsilon$ mildly biases the optimization procedure towards the original manifold, such that the loss function and gradients remain feasible to estimate with finite statistics. An alternate approach is to directly penalize distance between $\Re{\tilde{S}}$ and $S$ using a regularization term,
\begin{equation}
    \mathcal{L}_{\text{act}} \equiv \epsilon \frac{1}{Z} \int dx \, e^{-S(x)} \left| S(x) - \Re{\tilde{S}(x)} \right|.
\end{equation}
This term is minimized when $S = \Re{\tilde{S}}$, providing a bias towards remaining close to the original manifold. Though written as a path integral, this quantity can be estimated using the original samples, much like the main loss function and gradients.

Both of these regularizer terms were explored, however no severe overlap problem was observed during training when deformation parameters were restricted to those contained within the target Wilson loop observables. Final results are based on training without either term.

\bibliography{noise_refs}

\end{document}

%% file: contour1_cartoon.tex
\tikzset{every picture/.style={line width=0.75pt}} %set default line width to 0.75pt        

\begin{tikzpicture}[x=0.75pt,y=0.75pt,yscale=-1,xscale=1]
%uncomment if require: \path (0,310); %set diagram left start at 0, and has height of 310

%Shape: Axis 2D [id:dp25002381159098763] 
\draw  (15,115.58) -- (141.59,115.58)(27.66,12.6) -- (27.66,127.02) (134.59,110.58) -- (141.59,115.58) -- (134.59,120.58) (22.66,19.6) -- (27.66,12.6) -- (32.66,19.6)  ;
%Straight Lines [id:da012448710548571551] 
\draw [color={rgb, 255:red, 3; green, 10; blue, 167 }  ,draw opacity=1 ][line width=1.5]    (27.66,115.58) -- (116.81,115.58) ;
\draw [shift={(116.81,115.58)}, rotate = 0] [color={rgb, 255:red, 3; green, 10; blue, 167 }  ,draw opacity=1 ][fill={rgb, 255:red, 3; green, 10; blue, 167 }  ,fill opacity=1 ][line width=1.5]      (0, 0) circle [x radius= 3.48, y radius= 3.48]   ;
\draw [shift={(27.66,115.58)}, rotate = 0] [color={rgb, 255:red, 3; green, 10; blue, 167 }  ,draw opacity=1 ][fill={rgb, 255:red, 3; green, 10; blue, 167 }  ,fill opacity=1 ][line width=1.5]      (0, 0) circle [x radius= 3.48, y radius= 3.48]   ;
%Curve Lines [id:da22036725062847506] 
\draw [color={rgb, 255:red, 112; green, 178; blue, 63 }  ,draw opacity=1 ][line width=1.5]    (27.66,115.58) .. controls (36.47,108.71) and (50.34,92.65) .. (61.04,95.07) .. controls (71.74,97.5) and (62.19,56.64) .. (72.58,63.08) .. controls (82.97,69.52) and (92.41,72.35) .. (93.73,87.07) .. controls (95.05,101.79) and (78.57,125.26) .. (82.84,131.73) .. controls (87.1,138.21) and (112.47,118.96) .. (116.81,115.58) ;
\draw [shift={(116.81,115.58)}, rotate = 322.05] [color={rgb, 255:red, 112; green, 178; blue, 63 }  ,draw opacity=1 ][fill={rgb, 255:red, 112; green, 178; blue, 63 }  ,fill opacity=1 ][line width=1.5]      (0, 0) circle [x radius= 1.74, y radius= 1.74]   ;
\draw [shift={(27.66,115.58)}, rotate = 322.05] [color={rgb, 255:red, 112; green, 178; blue, 63 }  ,draw opacity=1 ][fill={rgb, 255:red, 112; green, 178; blue, 63 }  ,fill opacity=1 ][line width=1.5]      (0, 0) circle [x radius= 1.74, y radius= 1.74]   ;
%Curve Lines [id:da5448767767433973] 
\draw [color={rgb, 255:red, 208; green, 2; blue, 27 }  ,draw opacity=1 ][line width=1.5]  [dash pattern={on 5.63pt off 4.5pt}]  (27.71,83.07) .. controls (37.31,75.59) and (51.22,41.38) .. (62.97,44.42) .. controls (74.71,47.45) and (81.52,117.49) .. (95.01,137.73) .. controls (108.51,157.97) and (111.6,105.13) .. (116.81,101.07) ;
\draw [shift={(116.81,101.07)}, rotate = 322.05] [color={rgb, 255:red, 208; green, 2; blue, 27 }  ,draw opacity=1 ][fill={rgb, 255:red, 208; green, 2; blue, 27 }  ,fill opacity=1 ][line width=1.5]      (0, 0) circle [x radius= 2.61, y radius= 2.61]   ;
\draw [shift={(27.71,83.07)}, rotate = 322.05] [color={rgb, 255:red, 208; green, 2; blue, 27 }  ,draw opacity=1 ][fill={rgb, 255:red, 208; green, 2; blue, 27 }  ,fill opacity=1 ][line width=1.5]      (0, 0) circle [x radius= 2.61, y radius= 2.61]   ;
%Shape: Axis 2D [id:dp20850113919583269] 
\draw  (172,114.58) -- (298.59,114.58)(184.66,11.6) -- (184.66,126.02) (291.59,109.58) -- (298.59,114.58) -- (291.59,119.58) (179.66,18.6) -- (184.66,11.6) -- (189.66,18.6)  ;
%Straight Lines [id:da009943316285294879] 
\draw [color={rgb, 255:red, 3; green, 10; blue, 167 }  ,draw opacity=1 ][line width=1.5]    (184.66,114.58) -- (273.81,114.58) ;
\draw [shift={(273.81,114.58)}, rotate = 0] [color={rgb, 255:red, 3; green, 10; blue, 167 }  ,draw opacity=1 ][fill={rgb, 255:red, 3; green, 10; blue, 167 }  ,fill opacity=1 ][line width=1.5]      (0, 0) circle [x radius= 3.48, y radius= 3.48]   ;
\draw [shift={(184.66,114.58)}, rotate = 0] [color={rgb, 255:red, 3; green, 10; blue, 167 }  ,draw opacity=1 ][fill={rgb, 255:red, 3; green, 10; blue, 167 }  ,fill opacity=1 ][line width=1.5]      (0, 0) circle [x radius= 3.48, y radius= 3.48]   ;
%Curve Lines [id:da9273036543894462] 
\draw [color={rgb, 255:red, 112; green, 178; blue, 63 }  ,draw opacity=1 ][line width=1.5]    (184.66,114.58) .. controls (193.47,107.71) and (204.13,100.57) .. (214.83,103) .. controls (225.54,105.43) and (210.44,132.56) .. (220.83,139) .. controls (231.23,145.44) and (243.83,89) .. (248.83,100) .. controls (253.83,111) and (249.57,151.53) .. (253.83,158) .. controls (258.1,164.47) and (269.47,117.96) .. (273.81,114.58) ;
\draw [shift={(273.81,114.58)}, rotate = 322.05] [color={rgb, 255:red, 112; green, 178; blue, 63 }  ,draw opacity=1 ][fill={rgb, 255:red, 112; green, 178; blue, 63 }  ,fill opacity=1 ][line width=1.5]      (0, 0) circle [x radius= 1.74, y radius= 1.74]   ;
\draw [shift={(184.66,114.58)}, rotate = 322.05] [color={rgb, 255:red, 112; green, 178; blue, 63 }  ,draw opacity=1 ][fill={rgb, 255:red, 112; green, 178; blue, 63 }  ,fill opacity=1 ][line width=1.5]      (0, 0) circle [x radius= 1.74, y radius= 1.74]   ;
%Curve Lines [id:da4879760234687782] 
\draw [color={rgb, 255:red, 112; green, 178; blue, 63 }  ,draw opacity=1 ][line width=1.5]    (184.71,86.07) .. controls (194.31,78.59) and (210.83,64) .. (223.83,66) .. controls (236.83,68) and (205.83,77) .. (216.83,90) .. controls (227.83,103) and (234.83,86) .. (247.83,82) .. controls (260.83,78) and (271.94,88.33) .. (274.83,86.07) ;
\draw [shift={(274.83,86.07)}, rotate = 322.05] [color={rgb, 255:red, 112; green, 178; blue, 63 }  ,draw opacity=1 ][fill={rgb, 255:red, 112; green, 178; blue, 63 }  ,fill opacity=1 ][line width=1.5]      (0, 0) circle [x radius= 2.61, y radius= 2.61]   ;
\draw [shift={(184.71,86.07)}, rotate = 322.05] [color={rgb, 255:red, 112; green, 178; blue, 63 }  ,draw opacity=1 ][fill={rgb, 255:red, 112; green, 178; blue, 63 }  ,fill opacity=1 ][line width=1.5]      (0, 0) circle [x radius= 2.61, y radius= 2.61]   ;
%Straight Lines [id:da4922067466008355] 
\draw  [dash pattern={on 0.84pt off 2.51pt}]  (184.71,86.07) -- (274.83,86.07) ;
%Straight Lines [id:da19681082381448634] 
\draw [color={rgb, 255:red, 128; green, 128; blue, 128 }  ,draw opacity=1 ][line width=1.5]    (184.71,9.57) .. controls (186.38,11.24) and (186.38,12.9) .. (184.71,14.57) .. controls (183.04,16.24) and (183.04,17.9) .. (184.71,19.57) .. controls (186.38,21.24) and (186.38,22.9) .. (184.71,24.57) .. controls (183.04,26.24) and (183.04,27.9) .. (184.71,29.57) .. controls (186.38,31.24) and (186.38,32.9) .. (184.71,34.57) .. controls (183.04,36.24) and (183.04,37.9) .. (184.71,39.57) .. controls (186.38,41.24) and (186.38,42.9) .. (184.71,44.57) .. controls (183.04,46.24) and (183.04,47.9) .. (184.71,49.57) .. controls (186.38,51.24) and (186.38,52.9) .. (184.71,54.57) .. controls (183.04,56.24) and (183.04,57.9) .. (184.71,59.57) .. controls (186.38,61.24) and (186.38,62.9) .. (184.71,64.57) .. controls (183.04,66.24) and (183.04,67.9) .. (184.71,69.57) .. controls (186.38,71.24) and (186.38,72.9) .. (184.71,74.57) .. controls (183.04,76.24) and (183.04,77.9) .. (184.71,79.57) .. controls (186.38,81.24) and (186.38,82.9) .. (184.71,84.57) .. controls (183.04,86.24) and (183.04,87.9) .. (184.71,89.57) .. controls (186.38,91.24) and (186.38,92.9) .. (184.71,94.57) .. controls (183.04,96.24) and (183.04,97.9) .. (184.71,99.57) .. controls (186.38,101.24) and (186.38,102.9) .. (184.71,104.57) .. controls (183.04,106.24) and (183.04,107.9) .. (184.71,109.57) .. controls (186.38,111.24) and (186.38,112.9) .. (184.71,114.57) .. controls (183.04,116.24) and (183.04,117.9) .. (184.71,119.57) .. controls (186.38,121.24) and (186.38,122.9) .. (184.71,124.57) .. controls (183.04,126.24) and (183.04,127.9) .. (184.71,129.57) .. controls (186.38,131.24) and (186.38,132.9) .. (184.71,134.57) .. controls (183.04,136.24) and (183.04,137.9) .. (184.71,139.57) -- (184.71,140) -- (184.71,140) ;
%Straight Lines [id:da016350095710515156] 
\draw [color={rgb, 255:red, 128; green, 128; blue, 128 }  ,draw opacity=1 ][line width=1.5]    (274.71,9.57) .. controls (276.38,11.24) and (276.38,12.9) .. (274.71,14.57) .. controls (273.04,16.24) and (273.04,17.9) .. (274.71,19.57) .. controls (276.38,21.24) and (276.38,22.9) .. (274.71,24.57) .. controls (273.04,26.24) and (273.04,27.9) .. (274.71,29.57) .. controls (276.38,31.24) and (276.38,32.9) .. (274.71,34.57) .. controls (273.04,36.24) and (273.04,37.9) .. (274.71,39.57) .. controls (276.38,41.24) and (276.38,42.9) .. (274.71,44.57) .. controls (273.04,46.24) and (273.04,47.9) .. (274.71,49.57) .. controls (276.38,51.24) and (276.38,52.9) .. (274.71,54.57) .. controls (273.04,56.24) and (273.04,57.9) .. (274.71,59.57) .. controls (276.38,61.24) and (276.38,62.9) .. (274.71,64.57) .. controls (273.04,66.24) and (273.04,67.9) .. (274.71,69.57) .. controls (276.38,71.24) and (276.38,72.9) .. (274.71,74.57) .. controls (273.04,76.24) and (273.04,77.9) .. (274.71,79.57) .. controls (276.38,81.24) and (276.38,82.9) .. (274.71,84.57) .. controls (273.04,86.24) and (273.04,87.9) .. (274.71,89.57) .. controls (276.38,91.24) and (276.38,92.9) .. (274.71,94.57) .. controls (273.04,96.24) and (273.04,97.9) .. (274.71,99.57) .. controls (276.38,101.24) and (276.38,102.9) .. (274.71,104.57) .. controls (273.04,106.24) and (273.04,107.9) .. (274.71,109.57) .. controls (276.38,111.24) and (276.38,112.9) .. (274.71,114.57) .. controls (273.04,116.24) and (273.04,117.9) .. (274.71,119.57) .. controls (276.38,121.24) and (276.38,122.9) .. (274.71,124.57) .. controls (273.04,126.24) and (273.04,127.9) .. (274.71,129.57) .. controls (276.38,131.24) and (276.38,132.9) .. (274.71,134.57) .. controls (273.04,136.24) and (273.04,137.9) .. (274.71,139.57) -- (274.71,140) -- (274.71,140) ;
%Right Arrow [id:dp761231037087466] 
\draw  [draw opacity=0][fill={rgb, 255:red, 128; green, 128; blue, 128 }  ,fill opacity=1 ] (248.72,16.31) -- (267.13,30.52) -- (268.34,28.94) -- (272.43,37.14) -- (263.47,35.25) -- (264.69,33.68) -- (246.29,19.47) -- cycle ;
%Right Arrow [id:dp13272832472280505] 
\draw  [draw opacity=0][fill={rgb, 255:red, 128; green, 128; blue, 128 }  ,fill opacity=1 ] (210.55,19.84) -- (194.04,36.21) -- (195.45,37.63) -- (186.79,40.6) -- (189.83,31.97) -- (191.24,33.38) -- (207.74,17.01) -- cycle ;
%Curve Lines [id:da07870378678844014] 
\draw [color={rgb, 255:red, 74; green, 74; blue, 74 }  ,draw opacity=1 ]   (48.89,115.58) .. controls (49.73,109) and (47.91,104.23) .. (46.01,101.27) ;
\draw [shift={(45.83,101)}, rotate = 236.31] [color={rgb, 255:red, 74; green, 74; blue, 74 }  ,draw opacity=1 ][line width=0.75]      (0, 0) circle [x radius= 1.34, y radius= 1.34]   ;
\draw [shift={(48.7,108.1)}, rotate = 446.84] [fill={rgb, 255:red, 74; green, 74; blue, 74 }  ,fill opacity=1 ][line width=0.08]  [draw opacity=0] (5.36,-2.57) -- (0,0) -- (5.36,2.57) -- (3.56,0) -- cycle    ;
\draw [shift={(48.83,116)}, rotate = 278.13] [color={rgb, 255:red, 74; green, 74; blue, 74 }  ,draw opacity=1 ][line width=0.75]      (0, 0) circle [x radius= 1.34, y radius= 1.34]   ;
%Curve Lines [id:da7786738388766477] 
\draw [color={rgb, 255:red, 74; green, 74; blue, 74 }  ,draw opacity=1 ]   (57.5,114.64) .. controls (57.54,108.81) and (58.52,102.07) .. (60.93,95.38) ;
\draw [shift={(61.04,95.07)}, rotate = 290.15] [color={rgb, 255:red, 74; green, 74; blue, 74 }  ,draw opacity=1 ][line width=0.75]      (0, 0) circle [x radius= 1.34, y radius= 1.34]   ;
\draw [shift={(58.41,104.84)}, rotate = 457.22] [fill={rgb, 255:red, 74; green, 74; blue, 74 }  ,fill opacity=1 ][line width=0.08]  [draw opacity=0] (5.36,-2.57) -- (0,0) -- (5.36,2.57) -- (3.56,0) -- cycle    ;
\draw [shift={(57.5,115)}, rotate = 270] [color={rgb, 255:red, 74; green, 74; blue, 74 }  ,draw opacity=1 ][line width=0.75]      (0, 0) circle [x radius= 1.34, y radius= 1.34]   ;
%Curve Lines [id:da4539251046582702] 
\draw [color={rgb, 255:red, 74; green, 74; blue, 74 }  ,draw opacity=1 ]   (66.47,114.57) .. controls (66.06,105.35) and (75.56,102.9) .. (79.5,96) .. controls (83.46,89.07) and (87.42,81.16) .. (87.5,73.24) ;
\draw [shift={(87.5,73)}, rotate = 270] [color={rgb, 255:red, 74; green, 74; blue, 74 }  ,draw opacity=1 ][line width=0.75]      (0, 0) circle [x radius= 1.34, y radius= 1.34]   ;
\draw [shift={(71.44,104.33)}, rotate = 489.22] [fill={rgb, 255:red, 74; green, 74; blue, 74 }  ,fill opacity=1 ][line width=0.08]  [draw opacity=0] (5.36,-2.57) -- (0,0) -- (5.36,2.57) -- (3.56,0) -- cycle    ;
\draw [shift={(84.95,84.99)}, rotate = 474.23] [fill={rgb, 255:red, 74; green, 74; blue, 74 }  ,fill opacity=1 ][line width=0.08]  [draw opacity=0] (5.36,-2.57) -- (0,0) -- (5.36,2.57) -- (3.56,0) -- cycle    ;
\draw [shift={(66.5,115)}, rotate = 265.63] [color={rgb, 255:red, 74; green, 74; blue, 74 }  ,draw opacity=1 ][line width=0.75]      (0, 0) circle [x radius= 1.34, y radius= 1.34]   ;
%Curve Lines [id:da11305718845382184] 
\draw [color={rgb, 255:red, 74; green, 74; blue, 74 }  ,draw opacity=1 ]   (94.45,116.4) .. controls (93.43,124.89) and (90.59,126.99) .. (88.62,131.71) ;
\draw [shift={(88.5,132)}, rotate = 111.8] [color={rgb, 255:red, 74; green, 74; blue, 74 }  ,draw opacity=1 ][line width=0.75]      (0, 0) circle [x radius= 1.34, y radius= 1.34]   ;
\draw [shift={(92.4,124.38)}, rotate = 286.65999999999997] [fill={rgb, 255:red, 74; green, 74; blue, 74 }  ,fill opacity=1 ][line width=0.08]  [draw opacity=0] (5.36,-2.57) -- (0,0) -- (5.36,2.57) -- (3.56,0) -- cycle    ;
\draw [shift={(94.5,116)}, rotate = 96.34] [color={rgb, 255:red, 74; green, 74; blue, 74 }  ,draw opacity=1 ][line width=0.75]      (0, 0) circle [x radius= 1.34, y radius= 1.34]   ;
%Curve Lines [id:da9507709065370622] 
\draw [color={rgb, 255:red, 74; green, 74; blue, 74 }  ,draw opacity=1 ]   (100.5,116.35) .. controls (100.5,121.9) and (100.56,123.01) .. (103.28,125.78) ;
\draw [shift={(103.5,126)}, rotate = 45] [color={rgb, 255:red, 74; green, 74; blue, 74 }  ,draw opacity=1 ][line width=0.75]      (0, 0) circle [x radius= 1.34, y radius= 1.34]   ;
\draw [shift={(100.73,121.49)}, rotate = 267.40999999999997] [fill={rgb, 255:red, 74; green, 74; blue, 74 }  ,fill opacity=1 ][line width=0.08]  [draw opacity=0] (5.36,-2.57) -- (0,0) -- (5.36,2.57) -- (3.56,0) -- cycle    ;
\draw [shift={(100.5,116)}, rotate = 90] [color={rgb, 255:red, 74; green, 74; blue, 74 }  ,draw opacity=1 ][line width=0.75]      (0, 0) circle [x radius= 1.34, y radius= 1.34]   ;

% Text Node
\draw (40.76,120.51) node [anchor=north west][inner sep=0.75pt]    {$\theta$};
% Text Node
\draw (34.89,23.86) node [anchor=north west][inner sep=0.75pt]   [align=left] {invalid $\displaystyle \widetilde{\theta }( \theta )$};
% Text Node
\draw (91.37,59.52) node [anchor=north west][inner sep=0.75pt]   [align=left] {valid $\displaystyle \widetilde{\theta }( \theta)$};
% Text Node
\draw (198.76,117.51) node [anchor=north west][inner sep=0.75pt]    {$\phi$};
% Text Node
\draw (200.45,45.03) node [anchor=north west][inner sep=0.75pt]   [align=left] {valid $\displaystyle \widetilde{\phi }( \phi)$};
% Text Node
\draw (187.37,142.52) node [anchor=north west][inner sep=0.75pt]   [align=left] {valid $\displaystyle \widetilde{\phi }( \phi)$};
% Text Node
\draw (208,6) node [anchor=north west][inner sep=0.75pt]   [align=left] {identified};

\end{tikzpicture}